\documentclass[amsmath, amssymb,reprint,nofootinbib]{revtex4-1}
\usepackage{amscd}
\usepackage{amsmath}
\usepackage{amssymb}
\usepackage{amsthm}
\usepackage{dcolumn}
\usepackage{booktabs}
\usepackage{dashrule}
\usepackage{bm}
\usepackage[utf8]{inputenc}
\usepackage[T1]{fontenc}
\usepackage{float}
\usepackage{graphicx}
\usepackage{import}
\usepackage{mathtools}
\usepackage{natbib}
\floatstyle{plaintop}
\usepackage{subfiles}
\usepackage{tabularx}
\usepackage{verbatim}
\usepackage{xcolor}
\usepackage{subcaption}

\newcommand{\eqn}{\begin{equation}}
\newcommand{\equ}{\end{equation}}
\newcommand{\fig}{\begin{figure}}
\newcommand{\fiu}{\end{figure}}

\def\dashedrule#1#2#3{{
	\dimen1=#2 \divide\dimen1 by 2
	
	\def\@ruledash{%
		\rule{\dimen1}{0pt}%
		\rule[0.5ex]{#1}{0.4pt}%
		\rule{\dimen1}{0pt}}%
		
	\count1=0\loop%
	\ifnum\count1<#3%
		\advance\count1 by 1%
		\@ruledash%
	\repeat}}

\begin{document}
\title{Machine Learning for Materials Developments in Metals Additive Manufacturing}
\author{N.S. Johnson}
\affiliation{Alliance for the Development of Additive Processing Technologies, Colorado School of Mines, Golden, CO 80401}
\affiliation{Los Alamos National Laboratory, Los Alamos, NM  87544}

\author{P. S. Vulimiri}
\author{A. C. To}
\affiliation{Department of Mechanical Engineeering \& Materials Science, University of Pittsburgh, Pittsburgh, PA 15261}

\author{X. Zhang}
\author{C.A. Brice}
\affiliation{Alliance for the Development of Additive Processing Technologies, Colorado School of Mines, Golden, CO 80401}

\author{B.B. Kappes$^*$}
\affiliation{Alliance for the Development of Additive Processing Technologies, Colorado School of Mines, Golden, CO 80401}

\author{A.P. Stebner}
\affiliation{Alliance for the Development of Additive Processing Technologies, Colorado School of Mines, Golden, CO 80401}

\begin{abstract}
In metals additive manufacturing (AM), materials and components are concurrently made in a single process as layers of metal are fabricated on top of each other in the near-final topology required for the end-use product.
Consequently, tens to hundreds of materials and part design degrees of freedom must be simultaneously controlled and understood; hence, metals AM is a highly interdisciplinary technology that requires synchronized consideration of physics, chemistry, materials science, physical metallurgy, computer science, electrical engineering, and mechanical engineering.
The use of modern machine learning approaches to model these degrees of freedom can reduce the time and cost to elucidate the science of metals AM and to optimize the engineering of these complex, multidisciplinary processes.
New machine learning techniques are not needed for most metals AM development; those used in other sects of materials science will also work for AM.
Most prolifically, the density functional theory (DFT) community  has used many of them since the early 2000s for evaluating numerous combinations of elements and crystal structures to discover new materials.
This materials technologies-focused review introduces the basic mathematics and terminology of machine learning through the lens of metals AM, and then examines potential uses of machine learning to advance metals AM, highlighting the many parallels to previous efforts in materials science and manufacturing while also discussing new challenges and adaptations specific to metals AM. 
\end{abstract}

\maketitle
\tableofcontents

\vspace{-2em}
\section{Motivation}
\begin{figure*}[t]
	\includegraphics[width=1\linewidth]{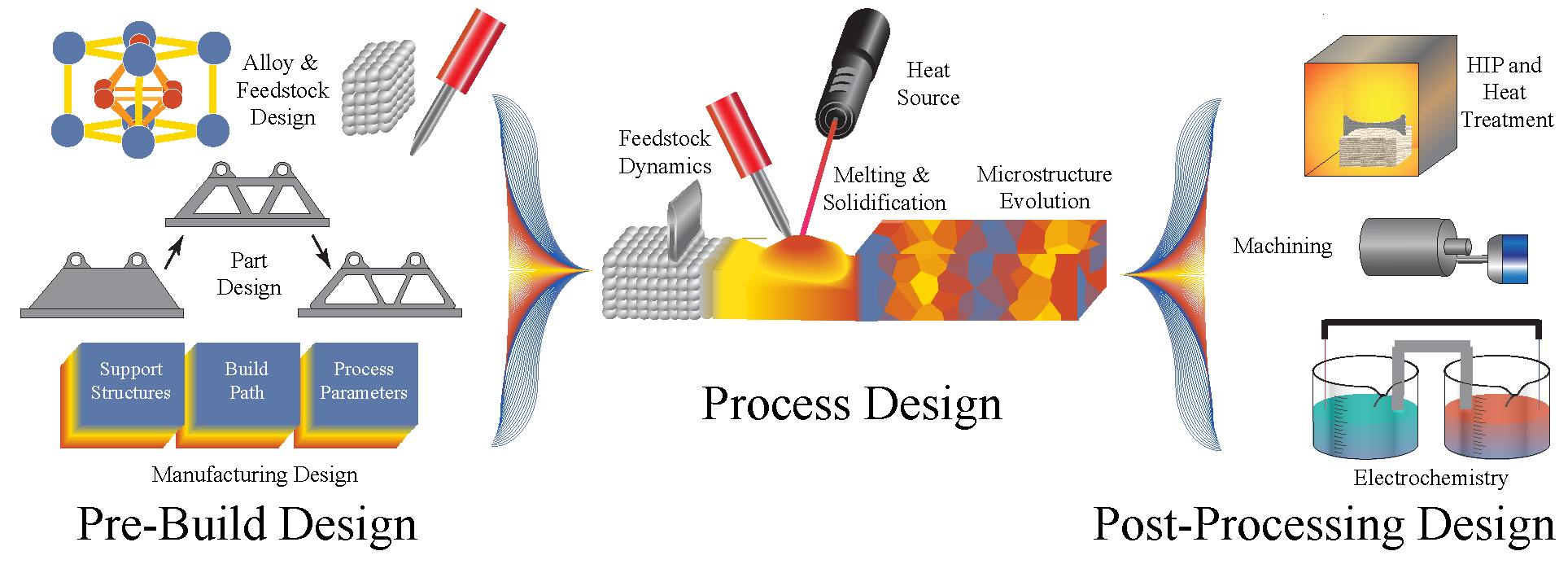}
	\caption{The design space of metals additive manufacturing spans many engineering disciplines since the material and part are made at the same time. As shown in this schematic, alloys, parts, and manufacturing process designs are concurrently considered in the "pre-build" phase. The physics of the process itself may then be modeled, including feedstock dynamics, thermodynamics and kinetics of melting, solidification, and thermal histories, which dictate the final microstructure. Today, post processing treatments are typically performed as secondary processes, though the future points to ``hybrid manufacturing" processes where they are also incorporated at the point of fabrication.}
	\label{AMgene}
\end{figure*}

Metals additive manufacturing (AM) has created a paradigm shift in they way metal components are manufactured; materials and parts are fabricated simultaneously using a single machine, highly complex geometries are possible, and local variations of microstructure-property relationships may be realized through local process variations. Although decades of scientific and engineering work in industry, academia, and government have resulted in the commercialization of metals AM technologies, the consistency and quality of parts and materials are still open challenges for many applications. In recent decades, Integrated Computational Materials Engineering (ICME) approaches have proven to accelerate the development and adoption of materials technologies \cite{Panchal2013}. Traditionally, ICME approaches incorporate physics-based experimental data with simulations that span different length and time scales. However, for metals AM, much of the physics are still being discovered; hence, the development of comprehensive, computationally feasible physics-first approaches to ICME are still an open challenge. The diverse array of promises and problems in AM has resulted in a field of study that is rich with data -- so much so that our ability to store and analyze the data is challenged. At the same time, this wealth of data is motivating a paradigm shift to incorporate machine learning into ICME approaches. 

\subsection{Background}
The 20th century saw the maturation of materials science and engineering as a field of study, enabling targeted materials discoveries and innovations for specific applications. Over the past several decades, materials development cycles have greatly accelerated by formulating materials problems through the process-structure-property-performance paradigm \cite{Olson2000, Panchal2013}. 

The process-structure-property-performance (PSPP) paradigm is a core philosophy in materials science and engineering that governs how the manufacturing of a material determines its ability to be used in different engineering applications. The PSPP relationships break down materials development into four key areas of scientific and engineering interest \cite{Olson2000}.

In AM, the \textbf{processing} of a material is dictated by the thermal, mechanical, and chemical changes experienced during its manufacture. Controllable machine parameters like energy density of the heat source, the path in which material is deposited or fused, the order in which part layers are manufactured, or the location of parts on the build plate are determining factors of the material process history. Table \ref{table:design_space} shows many of the controllable parameters common to laser-based additive manufacturing systems. The choice of these parameters largely impacts the processing history. The true processing history, however, is better described by the thermal history of the build volume, both during manufacture and post-processing, the mechanical forces it experiences, and any chemical reactions that occur in or on the part. Processing \textit{routes} are often discussed in AM and typically refer to beneficial or detrimental processing histories that impact the part's structure.

The \textbf{structure} of a material is a wide-ranging concept that spans many length scales. Structure can refer to the crystallographic structure at the atomic scale, to the morphology and orientation of grains at the mesoscale, to the geometry being manufactured at the macroscale. Microstructure is a term often used in materials science referring to a specific subset of the material structure. Microstructure for metals most commonly refers to grain and sub-grain level information like material phases, grain morphologies, texture, and any defects like pores or dislocations that might be present. Microstructures are often considered in analysis of material structures because they fundamentally dictate a material's properties. 

The \textbf{properties} of a material are characteristics that determine its qualities. Properties of metals AM parts that have been of interest are wide-ranging and they vary depending on the desired engineering application of the part. Mechanical properties are some of the most studied for AM metals since the majority of metals applications are structural. Other properties of interest include thermal conductivity, which determines the heat transfer through an AM part, chemical properties, like corrosion resistance, and optical properties, like reflectivity. 

The \textbf{performance} of a part is its ability to be successfully implemented in an engineering application. Performance can be viewed through the lifetime of an AM part when subjected to the mechanical, thermal, chemical, etc., forces it will experience. Early additively manufactured alloys showed degraded-to-comparable static properties compared to traditionally manufactured alloys \cite{Spierings2013}. Further research and development improved the static properties of AM materials, yet high microstructure variability and defect density can still cause AM material to fail unexpectedly in fatigue limited applications \cite{Wycisk2014, Edwards2014}. Some recent AM developments have resulted in material properties that exceed those of traditionally manufactured materials\cite{Probstle2016,Gallmeyer2019, Martin2017, Wang2017a, Liu2017a, Zhu2018}. Ultimately AM processes are unique relative to other metal fabrication techniques and it is difficult to make fair comparisons regarding performance across various manufacturing methods. When properly designed, AM parts can meet the intended performance needs in a wide variety of end-use applications. The large combinatorial space of manufacturing options in AM often obfuscates how proper design choices can be made.

The materials scientist interacts with the process-structure-property paradigm in traditionally manufactured materials. Traditional material manufacturing can be phrased in a cause-and-effect relationship between process, structure, and property. Once the material has been developed and characterized by the materials scientist or engineer, another engineer then considers the property-performance linkage. Since material is made separately from an engineered part in traditional manufacturing, the PSPP paradigm can be broken up into these two separate sets of relationships. In AM, the material and the part are made simultaneously. Simultaneous material-part manufacturing motivates consideration of linkages across the entire PSPP paradigm. The ICME approach to materials science is focused on modeling, bridging, and predicting relationships throughout the PSPP paradigm.

Computational materials science and engineering has enabled the prediction of microstructure from processing and of properties from microstructure, reducing the need for costly and time consuming experimentation in discovering or developing a new material and/or its manufacturing. Today, ICME approaches tightly integrate physics-based computational models into the industrial design process, allowing the desired performance requirements of a part to guide the design of a material. Alloy specific examples include low-RE Ni superalloys for better turbine performance \cite{Pollock2016} and lower cost and radioactive element free Ferrium S53 alloy designed for corrosion-resistant landing gears \cite{Olson2014}. Both cases reduced materials innovation timelines from decades to years, demonstrating the practical capability of designing and qualifying new materials within an industrial product development cycle. Generalizing and accelerating this capability across different industries and materials is a primary goal of the Materials Genome Initiative (MGI) \cite{MGI}.

Predicting PSPP linkages in metals AM is difficult with existing physics-based ICME approaches. The physics of AM processes are more complex than traditional fabrication methods, like casting, as they involve rapid solidification, vaporization and ingestion of volatile elements, and complex thermal history that consists of dozens of heating and cooling cycles, each one different. Furthermore, all of these additional complexities vary from one location to another within a part, and from part to part within a build volume. For AM, physics-based ICME tools have been mostly developed through attempts to adopt legacy manufacturing models to AM data, with some success. However, today's relatively low cost and time for performing AM processing experiments has led to metals AM development being largely combinatorial, with a chief strategy of adopting AM processing to legacy alloys that were developed for other types of manufacturing using extensive design of experiments.

It is with awareness of the large amounts of data being generated in AM through these combinatorial development cycles that machine learning (ML) has been targeted to accelerate AM innovations and their commercialization. Machine learning as a technology development accelerator has shown wide application in recent years across fields including finance \cite{Bose2001}, molecule design for genomics, chemistry and pharmacology \cite{Gomez-Bombarelli2018}, social networking \cite{Brusilovsky2007} and, most relevant to this review, materials science and engineering \cite{Wagner2016, Ramprasad2017, Butler2018}. Still, the use of ML in materials science was relatively limited for a variety of reasons, especially the lack of large curated datasets amenable to existing ML methodologies. Through the work done under the MGI, this data limitation was identified as a primary impediment to future materials innovations \cite{MGI}. In response, there has been significant recent investment in materials database developments to better enable materials data informatics innovations. It is now recognized and accepted that ML frameworks can couple legacy physics-based ICME tools with experimental data to produce more accurate process-structure-property models and to automate the iteration of designed experiments for model improvement and optimized materials \cite{Rajan2005, Agrawal2016, Butler2018, Ball2019, Druzgalski2020}.

We proceed to review how the paradigm shift from purely physics-based to coupled physics-based/data-driven ICME approaches can be made through solving metals AM challenges. We begin by phrasing terms and ideas from AM in ways that are compatible with machine learning. We provide a basic review of machine learning algorithms and how they can be applied to additive manufacturing. Following this introduction to using ML for AM problems, we review other uses of machine learning in materials science and engineering and state the uses of such approaches for solving AM challenges. 

\section{Phrasing Additive Manufacturing as a Machine Learning Problem}\label{phrasing}
While machine learning may seem abstract at first, it can be expressed and understood in plain terms. Many of the tenets and frameworks for machine learning are based in mathematical operations that are likely familiar to any scientist or engineer, but applied in new ways. In this section, we proceed to define the basic terminology and classes of machine learning and data. A list of machine learning algorithms used in the papers cited in this review can be found in Table \ref{ML}. The following section details general terminology and intuition for the application of AM. Specific ML algorithms are then introduced specific to contextual AM examples in Section III.

Machine learning algorithms are mathematical constructs that may be used as scientific and/or engineering tools when warranted. They are not appropriate for all science and engineering problems - just as finite element simulations should not be used to study the mechanics of discrete interfaces or single atomic bonds between two atoms or DFT should not be used to simulate mm-sized polycrystals, machine learning algorithms should not be used to model data that lack statistical correlations. Hence, the first questions every scientist and engineer considering the use of machine learning approaches should ask and answer is: "how are the data statistically distributed?", and "are there statistical correlations between the data features of interest?" Once this is complete, then a researcher can decide if ML is appropriate.

If the data lack clear statistical correlations using basic probability analyses, machine learning is not a "magic box" that can suddenly make such correlations evident. Similarly, if the statistical distributions of the data are featureless except for an occasional outlier, machine learning cannot meaningfully fit a model that is based on statistical distributions. 

Today, many scientists and engineers are embracing the approach that "we will machine learn it," without understanding how to evaluate if machine learning is an appropriate tool to apply to a problem or not. One unpublished example in AM of a problem that ML is not well suited for is building a model to predict the location of a maximum pore within a powder bed laser fusion build. A maximum pore is a statistical outlier - usually one of thousands-to-millions, depending on the size of the part being built. Even though the pore may occur in the exact same position of the build volume if the same part is built over-and-over again (i.e., it is highly repeatable), the fact that it is a statistical anomaly means that nearly all machine learning algorithms are built to ignore it. Once this understanding is at hand, then a researcher can decide if ML is appropriate.

Still, most data of interest in metals AM have strong statistical features, as we will proceed to discuss in more detail in the examples given in this review. Once some basic statistical analysis of the data of interest has been performed and it has been determined that there are quantifiable correlations between the inputs and outputs, or across different inputs, and that there are also statistical features that describe the distributions of the data, then a researcher can proceed to consider data featurization and processing, and then tune and evaluate the performances of machine learning models to find the best performers. We proceed to describe these techniques in more detail, after defining some basic terminology used in this article.

\begin{table*}[t]
    \renewcommand{\arraystretch}{0.8}
    \setlength{\tabcolsep}{5pt}
    \begin{center}
        \begin{tabular}{@{}llll@{}}
            \toprule
            \hline
             Parameter & range & step size & levels \\ \midrule
            \hline
            Power & 100-200 W & 10 W & 10 \\
            Scan speed & 500-1000 mm/s & 100 mm/s & 5 \\
            Spot size & 50-100 $\mu$m & 10 $\mu$m & 5 \\
            Energy density & 1-5 J/mm$^2$ & 1 J/mm$^2$ & 5 \\
            Sample Build Direction & 0-180$^\circ$ & 90$^\circ$  & 3 \\
            Amount of recycled powder & 0-100\% & 10\% & 10 \\
            Hatch spacing & 0.1-0.50 mm & 0.1 mm  & 5 \\
            \hline
            \bottomrule
        \end{tabular}
        \caption{A possible design space for laser powder bed fusion additive manufacturing. There are over $10^4$ possible combinations of machine inputs, based on the listed ranges and step sizes. Any possible combination of these parameters is a point in the design space.}
        \label{table:design_space}
    \end{center}
\end{table*}

\subsection{The Design Space of Additive Manufacturing}
The \textbf{design space} of metals AM is the set of all PSPP relationships. More specifically, the term `design space' will be used throughout this article in reference to the set of AM data that is used and calculated by machine learning algorithms. An example design space for laser powder bed fusion (LPBF) of metals, the most industrially prolific of current metals AM technologies, is graphically depicted in Figure \ref{AMgene}. A complementary example of a process design space of LPBF is given in Table \ref{table:design_space}. Observable process phenomena may link the manufacturing parameters to the resulting materials properties, hence they may also be used to augment the manufacturing parameters and material properties within the design space. Examples include melt pool morphology, temperature history, and cooling rates.

A single combination of process parameters, observed process phenomenon, measured material properties, and a part's performance can be considered as a \textit{coordinate}, or point, in the design space. Single coordinates, defined this way, can sometimes lead to a multitude of material properties due to latent variables, unforseen complications, and the stochasticity of the process. Explicit consideration of process phenomenon in the design space coordinate can be used to more accurately establish unique points within the design space. In summary, any part that is processed under a single set of conditions and is observed to have a set thermal history and set of material properties can be considered to be manufactured \textit{at that point} in the design space.

While the design space of AM is vast, data cannot always be given to machine learning algorithms `as is.' It is important to consider the sources of data in the design space and how they need to be changed or curated for use with ML.

\begin{turnpage}
\begin{table*}
\caption{Several of the most widely used machine learning algorithms in materials science. }
\label{ML}
\begin{tabular}{p{2.5cm}p{3cm}p{4cm}p{5.5cm}p{5.5cm}}\hline
\raggedright Class of Algorithm & Examples & Applications & Strengths & Constraints \\ \hline 
Weighted neighborhood clustering & \raggedright Decision trees, Random Forest, k-Nearest neighbor & \raggedright Regression, Classification, Clustering and similarity & \raggedright These algorithms are robust against uncertainty in data sets and can provide intuitive relationships between inputs and outputs. See Ref. \cite{Quinlan1986} for a primer on clustering. & They can be susceptible to classification bias toward descriptors with more data entries than others. \\ 

\dashedrule{.1em}{.1em}{325} & & & & \\

\raggedright Linear dimensionality reduction & \raggedright Principle component analysis (PCA), Support vector regression (SVR), Nonnegative Matrix Factorization (NMF) & \raggedright Experimental design, model dimensionality reduction, model or experimental input/output visualization, descriptor analysis, regression  & \raggedright This type of algorithm can produce orthogonal basis sets that reproduce the training data space. They can also provide quick and accurate regression analysis. For a primer on PCA specifically, see Ref. \cite{Bro2014}. & The relationships studied must be linear in nature, and these algorithms are susceptible to bias when descriptors are scaled differently. \\ 

\dashedrule{.1em}{.1em}{325} & & & & \\

\raggedright Nonlinear dimensionality reduction & \raggedright t-SNE, Kernel ridge regression, Multidimensional metric scaling &\raggedright  Experimental design, model dimensionality reduction, model or experimental input/output visualization, descriptor analysis, regression &\raggedright These algorithms are robust against nonlinear input/output relationships and can help visualize similarity in high dimensional relationships. For accessible examples, see Refs. \cite{Tenenbaum2000, Roweis2000}. & Interpretation of high dimensional similarity can be difficult; while these algorithms are useful for visualizing relationships interpreting the \textit{why} of the relationship found is difficult. Global relationships can also be lost when nonlinear dimensionality reduction results are projected onto lower-dimensional spaces.\\ 

\dashedrule{.1em}{.1em}{325} & & & & \\

Search algorithms & \raggedright Genetic algorithms (GA), Evolutionary algorithms & Alloy design (in conjunction with a material modeling approach), model optimization. topology optimization for AM &  \raggedright Search algorithms are intuitive for material properties that can be described geometrically, such as topology optimization for weight reduction. They are efficient at searching spaces with multiple local extrema, such as finding local maxima of quality in multidimensional design spaces. For a useful application of genetic algorithms to process characterization, see Ref \cite{Grefenstette1986}. &  These success of these algorithms are highly dependent upon selection and mutation criteria. \\ 

\dashedrule{.1em}{.1em}{325} & & & & \\

\raggedright Neural Networks \& Computer Vision & \raggedright Artifical Neural Networks, Convolutional Neural Networks (CNN), General Adversarial Networks (GAN) & \raggedright Classification, regression, feature identification and extraction in images, simulation of atomic potentials, transfer learning, in situ process monitoring, feedback and control & Neural networks have successfully modeled processing and image data; the research and development surrounding NNs is among the most mature of any type of machine learning algorithm. & Neural networks tend to require large training datasets, especially for image analysis applications; however, transfer learning approaches can adopt NNs to small datasets. \\ \hline
\end{tabular}
\end{table*}
\end{turnpage}

\subsection{Data Sources}

\begin{table*}[t]
	\renewcommand{\arraystretch}{1.25}
	\caption{Types and sources of data common in materials science and, specifically, additive manufacturing. The entries under each vary from a source of data -- like a characterization technique -- to the data itself -- like a single measured scalar value.}
		\begin{tabular}{p{2.75cm}p{2.75cm}p{2.75cm}p{2.75cm}p{2.75cm}p{2.75cm}} \hline
		Scalar & Time Series & Spectral & Images & Categorical & Spatial \\ \hline 
		\raggedright Ultimate tensile strength& Stress-strain curve  & X-ray diffraction & TEM & Composition & 3D Model and Slicing Path (e.g. STL file)\\
		Hardness & Temperature Gradient  & \raggedright X-ray Photospectroscopy  & SEM & Quality & Scan path \\
		
		Toughness & Pyrometry &  \raggedright X-ray Dispersive Spectroscopy & Optical Metallography  & Crystal structure &  Part Orientation in Build Chamber\\
		
		Fracture Strength &Thermography   & & \raggedright X-ray Computed Tomography & \raggedright Melt Pool Morphology & Crystallographic Texture\\
		Density & \raggedright Differential Thermogravimetric Analysis & & High Energy Diffraction Microscopy& & \\
		
		Solidification Velocity & Differential Scanning Calorimetry & & & &\\
		Cooling rate & Chemorheology& & & &\\
		Solidus/Liquidus Temperature & Magnetometry& & & &\\
		\raggedright Enthalpy of Formation/Melting & & & & &\\ 
		Pore size &  & & & & \\
		Fatigue Properties & & & & &\\ \hline
		\end{tabular}
		
	\label{sources}
\end{table*}

Data, as a materials scientist normally thinks about the term, encompasses a vast range of sources and formats. Some of the most common sources of data used by materials scientists for AM can be seen in Table \ref{sources}. 

The most obvious data that materials scientists interact with are scalar values like modulus, ultimate tensile strength, laser scan speed, laser energy, layer height, etc. Distributions of scalars are also used such as grain size distribution or particle size distribution of AM feedstock. Many materials scientists interact with series data that can be subdivided into several more categories. Times series data can include a temperature measurement from a thermocouple during an AM build. Other series data include X-ray diffraction histograms or X-ray fluorescence spectra. 

Data can also take non-quantitative forms, often referred to as categorical data. These can include crystallographic structure, grain morphology, or the shape of an AM part. In many cases, these categorizations can be converted into quantitative data by measuring a feature such as the major and minor axis length of a grain. More difficult to quantify categorical data in AM includes melt pool morphology and track solidification defects like ``balling" or ``lack of fusion/delamination."

Images are some of the most commonly obtained data sources in materials science and are taken from a wide range of techniques. Light optical microscopy, scanning electron microscopy, and transmission electron microscopy images are all collected to study material structure. Materials processing images may include computed tomography radiographs and/or 3D reconstruction of a melt pool and thermal measurements using two-color pyrometry. Images can be treated as a data point on their own, but they are often analyzed to extract other data such as measuring grain size from light optical microscopy or categorizing crystal structure from a transmission diffraction pattern. 

Data can also be esoteric, depending upon the problems within AM that are being addressed. For example, a vector field of particle flow from a computational fluid dynamics simulation can be considered data. The orientation distribution function of the material's texture can also be considered data. The 3D model and slicing path used to generate an AM part can be considered data. Limitations on what constitutes ``data'' in a materials science problem are not worth defining. Rather, it is more important to consider how data can be featurized for use with an ML algorithm, as this ability determines whether or not data are amenable to use for machine learning approaches.
\subsection{The Featurization and Curation of AM Data}\label{feat}

Featurization involves extracting information from a data set such that a machine learning algorithm can interpret relationships between features themselves or between features and desired processing outcomes like mechanical strength, surface roughness, shape, etc. The preprocessing step of featurizing data is crucial for successful implementation of machine learning algorithms. Improper featurization of data can impact prediction and classification errors \cite{Murdock2020}. 

Scalar data are possibly the easiest to work with because they are features themselves; scalar values are also often referred to as \textit{descriptors} in this context. Therefore, they do not necessarily require featurization but rather curation and organization for use with ML. In many cases, scientific and engineering studies of AM map scalar data related to machine parameters -- like those in Table \ref{table:design_space} -- to scalar measurements of material properties, like strength, modulus, surface roughness, density, etc. 

It is important that machine learning models are trained on datasets with a certain volume of data collected i.e. datasets with statistical variability -- in some cases this could be individual measurements of machine parameters and material properties distributed across the design space. In other cases, having repeated measurements at the same place in the design space can reveal variability in the manufacturing process. Scalar data can be featurized by conducting simple statistical tests to understand relationships that are already present in the dataset. Statistical information such as
\begin{itemize}
	\item Mean, median, and mode
	\item Standard deviation 
	\item The presence of outliers
	\item Correlation coefficients between parameters
	\item Type of distribution (Gaussian, Lorentzian, Weibull, etc.)
\end{itemize}
are fairly straightforward to assess. Understanding the basic statistical nature of the machine learning algorithms can prevent problems in the application of ML to AM. For example, heavily correlated inputs in a dataset impact the results of machine learning models. In the worst case scenario, having correlated inputs degrades the predictive capability of the algorithm being implemented; in the best case, it has no effect on model performance but slows down modeling and computation time by adding unnecessary computations  \cite{Li2018}. 

The type of distribution that best describes the data may guide the underlying assumptions that some machine learning models make. For example, as further discussed in Section III, Gaussian Process Regression assumes that a Gaussian distribution best describes the statistical variance of the data being modeled \cite{Ripley1981, Schabenberger2004}.

The removal of statistically correlated inputs (if necessary) combined with determining the statistical nature of the dataset is the curation of data.

Once it is determined that the data are properly featurized and curated, the next step is data organization. Collections of scalar values can be represented by several different mathematical tools before use with machine learning. Matrices are used in machine learning to represent multiple observations, typically stored in rows, of a set of features (columns), for example,
\begin{equation}
	\mathbf{X} = \begin{bmatrix} x_{1,1} & x_{1,2} & \hdots & x_{1,m} \\
						x_{2,1} & x_{2,2} & \hdots & x_{2,m} \\
						\vdots & \vdots & \vdots & \vdots \\
						x_{n,1} & x_{n,2} & \hdots & x_{n,m} \\
				\end{bmatrix}
	\label{matrix}
\end{equation}
where the columns of $\mathbf{X}$ represent the different features, out of $m$, and there have been $n$ repeated measurements of each. In some machine learning uses, the investigator wants to learn trends within the dataset. When varying dozens of parameters at a time, which is often the case in additive manufacturing, trends across multiple print parameters are not always obvious. In this case, a dataset of observations can be formatted into a matrix like in Eqn. \ref{matrix}. This is referred to as \textit{unlabeled} data. In other cases, the experimenter wants a predictive tool that allows them to ask: \textit{if} I print at \textit{these specific conditions}, what will be the result? In these cases, it is better to store the print parameters in a format like Eqn. \ref{matrix}, but have the resulting properties stored in a separate vector object, like
\begin{equation}
	\mathbf{Y} = \begin{bmatrix} y_{1} \\
				y_{2} \\
				\vdots \\
				y_{n} \\
				\end{bmatrix}.
			\label{outputs}
\end{equation}
In this case, the data has been separated into inputs and outputs. This is referred to as \textit{labeled} data.

A time series signal can be represented as a list of scalar values that are correlated in the time dimension. Indeed, it is possible to represent collections of time series signals using the mathematical form in Eqn. \ref{matrix}, where each column is a time step and each row is a different measurement. For some applications, this data processing approach will result in unnecessary data being used for modeling. For example, if looking for indications of defect formation, much of the collected data can be ignored. It can be reasonably assumed that defect formation occurs when certain signals change from an expected mean value, like a rapid rise in temperature or energy density of the laser. In these cases, it is better to search for indications of these changes away from the expected mean value instead of using the entire signal.

Featurization in this case is searching for aspects of the series that are correlated with a desired process outcome. For a time series signal, useful features include the signal maximum, minimum, locations with sharp changes in curvature, sudden changes in absolute value, and more. For other series data, such as diffraction histograms or spectral data, other features need to be considered. For diffraction histograms values like peak position, peak breadth, peak intensity, etc., are useful. Values measured in between peaks (i.e. the background noise) can likely be ignored. One of the major benefits of featurizing series data is removing unnecessary values -- indeed, the field of compressive sensing is focused around removing redundant information from series signals \cite{Candes2008}. This type of featurization is useful for formatting extracted scalar values into matrix and vector objects like Eqns. \ref{matrix} \& \ref{outputs}. Once features have been extracted from the series signal they can be represented as a collection of scalars. From there, the collection of features should be treated to the same statistical litmus tests described above for scalar values. It is worth noting that some machine learning algorithms use entire series for inputs and featurize them as part of the algorithm \cite{Long2007, Long2009, Kusne2015a}.

Featurization of images is an active area of research in computer vision, a sub field of machine learning. Images are characterized by a spatial correlation in intensity: discrete changes in intensity dividing regions/domains of comparable, or slowly varying, intensity. Images are also most often represented as matrices of spatially-correlated intensity. The image processing algorithms discussed elsewhere in this review rely on a matrix representation of images. There are many toolboxes available, both free and commercial, which can pre-process images for use in machine learning algorithms; for example, the MATLAB Computer Vision Toolbox and the C++/Python OpenCV libraries.

Featurization of images occurs in a wide variety of ways and will be discussed in-depth in Section III. However, it is worthwhile here to discuss \textit{filters}, one of the most common ways of extracting features from an image. A filter is a mathematical operation applied to a region of an image that changes or enhances that image. Filters can be used to remove noise or distortion from an image, blur the image, sharpen edges, and more. 

Filtering an image is computing the product of a matrix $\mathbf{w}$ with a matrix $\mathbf{f}(x,y)$. The function $\mathbf{f}$ is the pixel value of an image $I$ at location $(x,y)$. The filter is applied as the product
\begin{equation}
		\mathbf{g}(x,y) = \sum_{s = -m}^m \sum_{t=-n}^n \mathbf{w}(s,t) \mathbf{f}(x+s,y+t)
	\label{filter}
\end{equation}
where $\mathbf{g}(x,y)$ is the matrix resulting from the operation \footnote{The product between $\mathbf{w}$ and $\mathbf{f}$ is not valid in all locations due to mismatches in indices at the border of the image. There are special cases defined where either the weight matrix or the image needs to be modified; Szeliski \cite{Szeliski2011} and MATLAB's documentation \cite{MATLABPad} provide more information.}. The product in Eqn. \ref{filter} is a \textit{convolution} of $\mathbf{w}$ and $\mathbf{f}$. In certain cases a correlation operation is applied instead. More information on these operations can be found in the work of Szeliski et al. \cite{Szeliski2011}, or any online open resource discussing image filtering.

The product of filtering is another image $\mathbf{g}(x,y)$ that has been modified or enhanced to reveal aspects of the original image. The application of filters can identify edges, reveal bright spots, reduce noise, blur, and do more to an image. The filtered matrix $\mathbf{g}$ can also be used sn input for a machine learning application, like regression. Some machine learning applications like convolutional neural networks (CNNs) actually learn filters $\mathbf{w}$ themselves that maximize prediction accuracy in regression applications. 

While filters are perhaps the most common featurization tool for images, other featurization methods exist. N-point correlation functions have found extensive use in extracting features from materials microstructure data \cite{Fullwood2008}. 

Many machine learning algorithms can operate directly on machine inputs and processing outputs; however, it can be equally useful to measure the relationship between data points instead of the values of the data points themselves, i.e. using the covariance. The covariance is measured between two data points $\kappa (\mathbf{x},\mathbf{x}')$, instead of being a property of a single data point. The function $\kappa(\cdot,\cdot)$ is called a kernel function. The covariance between data points encodes cross-correlated information within the design space. Kernel functions can be used to assess the similarity of design space coordinates or to transform the feature space--for example, from a linear to a logarithmic space, or from a continuous to a logistic space--to better suit the underlying physics of the feature--target relationship \cite{KernelMethod}. Ways of calculating covariance are many and varied and will be defined explicitly throughout this review as they are used.

Once data has been pre-processed and featurized it can be used in a machine learning algorithm, but first it is collected it is important to consider some of the underlying assumptions of machine learning and ensure that the dataset being used meets those assumptions.
\subsection{The Assumptions Behind Machine Learning}
Two fundamental assumptions underpin the use of machine learning:
\begin{enumerate}
\item \textit{The Relational Hypothesis}: A correlative relationship exists between the data input to the ML model and the response of the system.
\item \textit{The Similarity Hypothesis}: Similar points in the design space will have similar properties.
\end{enumerate}

The relational hypothesis is a foundation for predictive models: after all, no prediction is possible in the absence of a correlative relationship between input and response.
 
The similarity hypothesis supposes that data are comparable: that according to some measure of similarity, similar input will produce similar output. 

There are two types of machine learning covered in this review: unsupervised and supervised. Unsupervised learning will find trends in a dataset that are indicative of the underlying behavior. Supervised learning will learn a function $f(\mathbf{x}) = y$ that encodes part of the PSPP relationship. We proceed to walk through toy examples of each type; keep in mind that these are simplified examples meant to provide intuition behind the uses of machine learning. Scientists and engineers should research machine learning models, their uses, and their specific underlying assumptions before applying them.

\subsection{Unsupervised Machine Learning}\label{unsupervised}

Unsupervised machine learning algorithms are used to identify similarities or draw conclusions from unlabeled data by relying on the similarity hypothesis. Unsupervised approaches are useful for visualizing or finding trends in high dimensional data sets, screening out irrelevant modeling inputs, or finding manufacturing conditions that produce similar material properties. 

Consider an experiment that varies three different manufacturing inputs $x_1, x_2, x_3$ and measures a single material property $y$. A distance metric can be defined between data points in the design space. For example, data can be collected at two points $\mathbf{a} = (x_{1}, x_{2}, x_{3})$ and $\mathbf{b} = (x_{1} + \delta, x_{2}, x_{3})$. The $\ell _2$ norm of $\mathbf{a}-\mathbf{b}$ yields

\eqn
|| \mathbf{a} - \mathbf{b}||_2 = \delta.
\equ

The value and magnitude of $\delta$ gives an inclination about how similar $\mathbf{a}$ and $\mathbf{b}$ are.
If $\delta$ is close to zero, then a researcher can say that $\mathbf{a}$ and $\mathbf{b}$ are similar.
As $\delta$ becomes larger a researcher can say $\mathbf{a}$ and $\mathbf{b}$ become more dissimilar.
The concept of `similar' manufacturing conditions may be easy to assess by an experimentalist when tuning only a few parameters at a time.
When taking into consideration tens or hundreds of design criteria, sometimes with correlated inputs, elucidating similar manufacturing conditions becomes difficult.
This vector distance approach is a simple, yet effective first glance at similarity in a design space and is generalizable to $n$ many design criteria.

Let us say that $\delta$ is small and that $\mathbf{a}$ and $\mathbf{b}$ are similar manufacturing conditions.
Now, consider a third point in the design space $\mathbf{c} = (x_{1} + \delta, x_2 + \delta, x_3)$ that has not yet been measured.
Since $\mathbf{c}$ was manufactured at similar conditions to $\mathbf{a}$, as measured by $||\mathbf{c} - \mathbf{a}||_2 = \sqrt{2}\delta$, then we may say that $\mathbf{a}$, $\mathbf{b}$, and $\mathbf{c}$ are all similar to each other. If the similarity hypothesis is correct then manufacturing with conditions $\mathbf{a}$, $\mathbf{b}$ and $\mathbf{c}$ should yield similar measurements of $y$.

To better understand why unsupervised learning is desirable for AM R\&D consider a research project with initial manufacturing inputs $\mathbf{a}$, $\mathbf{b}$, $\mathbf{c}$, $\mathbf{d}$, etc., and associated property measurements that have been tested. Measuring the remainder of all possible design space coordinates to map the process-structure-property-performance relationship quickly becomes prohibitive. Instead, researchers can use similarity metrics to determine whether or not a future test is worth running. Comparing the manufacturing inputs through vector distance gives a rough idea of the possible outcome before spending time and resources on running a test. If the intent is exploring design spaces then manufacturing at conditions \textit{furthest away} from previously observed points may be the answer. If looking for local maxima of quality, an operator would want to manufacture at conditions \textit{nearest to} the conditions currently known to have high quality.

Another common application of unsupervised learning is finding clusters in data sets that produce useful partitions of material behavior. Using vector distances as metrics of similarities can produce results that are analogous to creating process maps \cite{Beuth2001}, which is further discussed in Section \ref{viz}. 


The following demonstration of unsupervised learning is based on $k$-means clustering, a commonly used unsupervised machine learning clustering algorithm.

A researcher has acquired the datasets in Eqn. \ref{matrix} and wants to partition $x_j \in \mathbf{X}$ into groupings of print parameters that produce similar results.
However, there are several values of $x_j \in \mathbf{X}$ that lie between two extremes and the cutoff for `similar conditions' is not obvious.
Similarity metrics can be used to find demarcations in the dataset that indicate regions of similarity.
To begin, the data set is partitioned randomly into two groups, $\mathbf{X}_1$ and $\mathbf{X}_2$.
The centroids $m_1$, $m_2$ (or centers of mass, in engineering) of each grouping can be calculated as

\eqn
	\begin{split}
		m_1 & = \frac{1}{|\mathbf{X}_1|} \sum_{x_j \in \mathbf{X}_1} x_j \\
		m_2 & = \frac{1}{|\mathbf{X}_2|} \sum_{x_j \in \mathbf{X}_2} x_j. \\
		\label{moment}
	\end{split}
\equ
where $\left|\mathbf{X}\right|$ is the mean value of a grouping.
The measurements were randomly partitioned at first; the goal is to re-partition each set so that similar measurements are in the same set.
To do this, we can re-assign each set by

\eqn
	\begin{split}
		\mathbf{X}_1 & = \{x_i : ||x_i - m_1||_2 \leq ||x_i - m_2||_2 \} \\
		\mathbf{X}_2 & = \{x_j : ||x_j - m_2||_2 \leq ||x_j - m_1||_2 \}. \\
	\end{split}
	\label{reassign}
\equ

The re-assignment in Eqn. \ref{reassign} can be interpreted physically: if a measurement initially assigned to set $\mathbf{X}_1$ is closer in distance to the centroid of $\mathbf{X}_2$ then it is \textit{more similar} to the other set.
Thus, it is re-assigned.
Measuring the similarity of each data point to the mean of the groupings re-classifies these outliers into groupings that are more reflective of the position in the high dimensional design space, giving manufacturing designers insight into how design parameters are distributed in that space.

Once re-assignment is complete the centroids in Eqn. \ref{moment} can be re-calculated and updated.
Then, data points are re-assigned once more based on how similar they are to the centroid of each partition.
If the input settings $(x_1, x_2, x_3)$ are partitioned along with their corresponding measurements, then we have lists of input settings that are likely to give good/bad quality parts.
Further analysis can also be conducted, such as analyzing which regimes of inputs lead to good or bad quality - this is precisely what process maps represent.
The difference in this case is that $n$ many manufacturing conditions can be related to a quality metric simultaneously, with little to no human inspection or intervention.
Additionally, a researcher can dig further and analyze \textit{why} groups of input settings result in given quality for a material property.
\subsection{Supervised Machine Learning}
In a \textit{supervised machine learning algorithm} the goal is to determine a functional relationship $f(\mathbf{x}) = \mathbf{y}$ based on previous measurements of $\mathbf{y}$ at points $\mathbf{x}$ in the design space. That is, supervised machine learning algorithms relate manufacturing inputs to labeled output data. 

Functional mappings of input data $\mathbf{x}$ to process outcomes $\mathbf{y}$ can take the form of either regression or classification. In a regression problem, the goal is to find mappings between inputs $\mathbf{x}$ to continuous values of $\mathbf{y}$. An example includes predicting mechanical strength from processing conditions, where the process conditions can be continuous or discrete, like those in Table \ref{table:design_space}, and the output $\mathbf{y}$ can be any reasonable value of strength. A classification problem sorts inputs $\mathbf{x}$ into categories with associated labels. These classifications can be binary or one-of-many classes. An example would be training an algorithm to answer the question ``Will the build fail?" based on processing inputs, with the possible class labels being ``Yes" or ``No."

Functional relationships can take many forms, depending on the specific supervised ML algorithm being used. One method is to model the relationships as a vector product
\eqn
\mathbf{X}\beta = \mathbf{Y}.
\label{map}
\equ
where $\beta$ is a vector of coefficients that weight the machine inputs to approximate an entry in $\mathbf{Y}$. 

A researcher usually seeks this relationship through the measurements they have observed; in this case, the measurements are stored in the matrices of Eqns. \ref{matrix} \& \ref{outputs}.
A common method to find a vector representation of $\beta$, and a critical element in most machine learning algorithms, is through least squares regression. Least squares regression finds $\beta$ through a minimization problem, given by
\eqn
\min || \mathbf{X}\beta - \mathbf{Y} ||_{2}^{2}.
\label{leastsquares}
\equ
Equation \ref{leastsquares} can be interpreted analogously to similarity measurements for unsupervised algorithms: the closer that $\mathbf{X}\beta - \mathbf{Y}$ is to zero, the more similar $\mathbf{X}\beta$ is to $f(\mathbf{x})$.

The methods of solving equation \ref{leastsquares} are many and varied; indeed, much of this review will focus on finding solutions to Eqn. \ref{leastsquares} for various problems throughout additive manufacturing.
The result is an approximation to the functional relationship $f(\mathbf{x}) = \mathbf{y}$.
A new point of interest in the design space $\mathbf{x'}$ can be chosen and its associated material property $\mathbf{y'}$ can be predicted by computing

\eqn
\mathbf{x'}\beta = \mathbf{y'}.
\equ
This simple example demonstrates how functional relationships can elucidate more information about design spaces from previously generated data.

\subsection{Error Metrics}\label{errormetrics}
Models that are used to predict values, whether numerical regression or classification algorithms, must have metrics to assess success. There are a multitude of error metrics that are used in the machine learning community. Different error metrics provide different information about the model, such as its ability to predict mean values, its robustness against outliers, and uncertainty in predictions, amongst other information. Many different error metrics have been formulated by the statistics community and used by the ML community \cite{Navidi2006}. Here, we review many of the most commonly used error metrics. For readers interested in more in depth discussion and examples, the website DataQuest provides an open source article about common error metrics\cite{DataQuestError}, explanations of commonly used error metrics and their benefits/drawbacks can be found in Table V of Shan et al\cite{Shan2010}, and Botchkarev wrote a review article detailing different error metrics used by the machine learning community over time\cite{Botchkarev}.

The following parameter definitions are used in the ensuing basic introduction of common error metrics and the remainder of the article.
\begin{itemize}
	\item $\hat y$ -- the value predicted by a regression algorithm $f(\mathbf{x}) = \hat y$
	\item $y$ -- the actual value of a material process/structure/property at input location $\mathbf{x}$
	\item $n$ -- the sample size used to train a machine learning algorithm
\end{itemize}

The mean absolute error (MAE) assesses the absolute residual between the predicted value of a regression problem and the actual value. It is calculated as the absolute difference between predicted value $\hat y$ and the actual value $y$, normalized by the sample size. Stated mathematically, the mean absolute error is 
\begin{equation}
	\text{MAE} = \frac{1}{n} \sum_{i=1}^n \left|y_i - \hat y_i \right|.
	\label{MAE}
\end{equation}
MAE penalizes error linearly. The MAE penalizes outliers in the data with the same magnitude as data points lying close to the mean. The mean absolute error can also be changed into a percentage, the mean absolute percentage error (MAPE) by normalizing each individual error measurement against the actual value $y$. Stated mathematically, 
\begin{equation}
	\text{MAPE} = 100 \times \frac{1}{n} \sum_{i=1}^n \left|\frac{y_i - \hat y_i}{y_i}\right|.
	\label{MAPE}
\end{equation}

Other error metrics highlight the impact of outliers on the dataset. The mean squared error (MSE) squares the difference term in Eqn. \ref{MAE} to produce
\begin{equation}
	\text{MSE} = \frac{1}{n} \sum_{i=1}^n |y_i - \hat y_i|^2.
	\label{MSE}
\end{equation}
The MSE penalizes error quadratically. Outliers in the dataset will have a much larger impact on MSE than they will on the MAE. A downside of the MSE is that the errors are reported as the square of the units being predicted by the model. Some users wish to have an error with the same units as the value being predicted; thus the root mean squared error (RMSE) adds a square root such that
\begin{equation}
	\text{RMSE} = \sqrt{\frac{1}{n} \sum_{i=1}^n |y_i - \hat y_i|^2}.
	\label{RMSE}
\end{equation}

All of the above metrics produce a measure of the absolute value of the error in the model. In some cases it is useful to know if a model is \textit{over} predicting the value (negative error) or \textit{under} predicting the value (positive error). In these cases, the MAE can be modified to the mean percentage error (MPE), given as
\begin{equation}
	\text{MPE} = \frac{1}{n} \sum_{i=1}^n \left( \frac{y_i - \hat y_i}{y_i}\right).
	\label{MPE}
\end{equation}
The MPE can reveal if a machine learning prediction algorithm is skewed towards certain types of values.

\begin{figure}
	\includegraphics[width=1\linewidth]{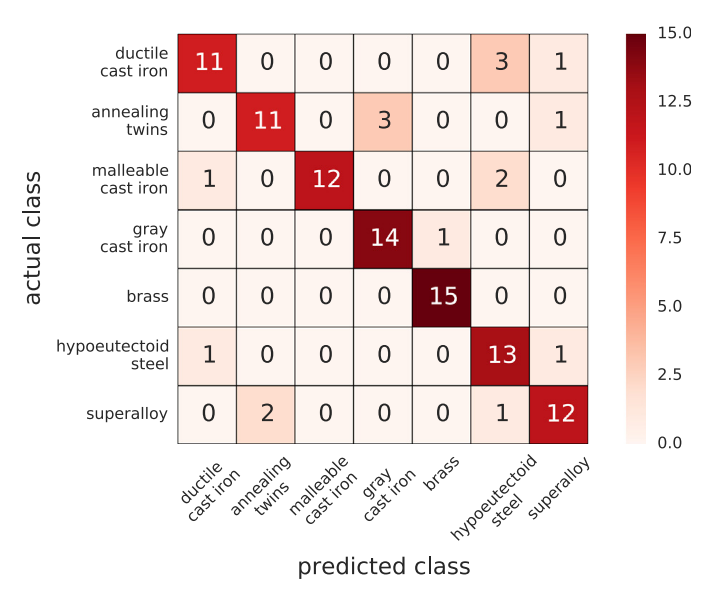}
	\caption{A confusion matrix used in a study by DeCost and Holm \cite{DeCost2015}. The goal of the study was to classify materials based on images of their microstructures. The main diagonal of the matrix represents correct classifications. In the case of the upper-leftmost entry, 11 images of ductile cast iron were correctly identified as ductile cast iron. The upper-rightmost entry indicates that 1 image of ductile cast iron was incorrectly classified as a superalloy.}
	\label{confusionmatrix}
\end{figure}

All of the above error metrics are suitable for regression problems with continuous value of $\hat y$. In the case of a classification problem, where the outputs are non-numerical, a non-numerical method of measuring error must be defined. While several methods have been developed \cite{Metrics2018, Metrics2019} a common method is to use a \textit{confusion matrix}. A confusion matrix displays the percentage of classifications that were correctly identified, as well as the percentage of classifications made to the wrong class. An example confusion matrix can be seen in Figure \ref{confusionmatrix}. In this figure, the main diagonal of the figure displays the percentage of data points that were correctly identified by class. The off-diagonal components display when a certain class was mis-identified as another class and how often it occurred.

\subsection{The Bias-Variance Tradeoff and Model Validation}\label{bvar}

\begin{figure*}
	\includegraphics[width=0.75\linewidth]{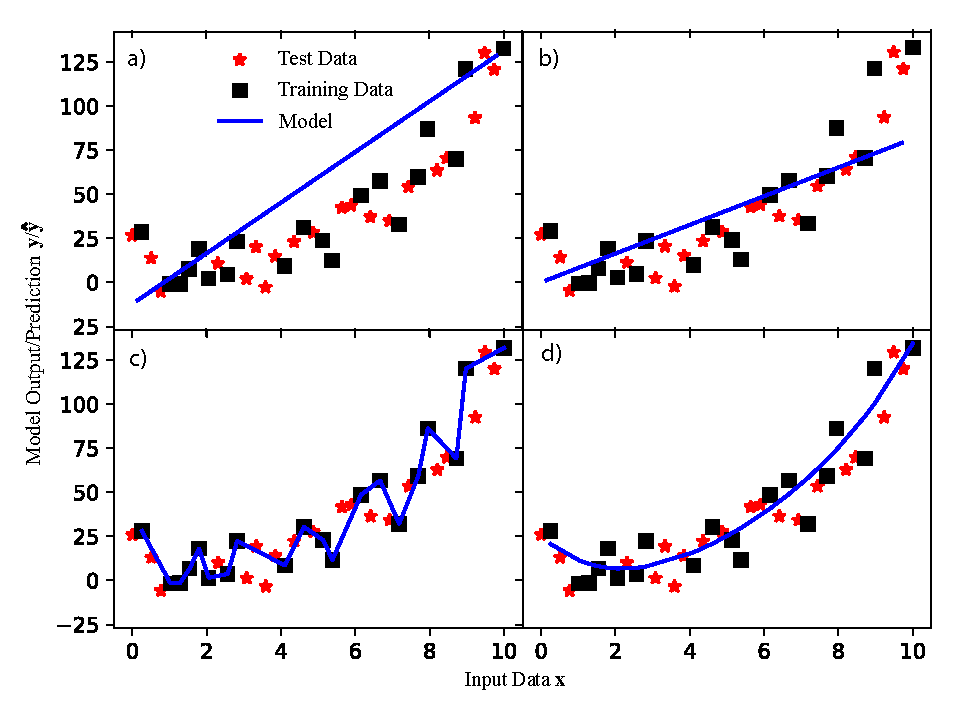}
	\caption{Illustrations of high bias and high variance models. A toy dataset was generated from the polynomial $y=5+0.1x+0.1x^2+0.1x^3+0.002x^4+ \text{Random Noise}$. The fits in a) and b) are both parameterizations of a model. Each model (line) in both fits has approximately the same error but does not accurately capture the behavior of the data due to poor model assumptions; in this case, fitting a first order polynomial to a dataset generated from a fourth order polynomial. This is an example of high bias models. A twentieth order polynomial was fit to a subset of the full dataset in c), shown in blue. While the model has very good predictive error for the training dataset it will not extrapolate well to the data in the testing set; this is overfitting or high variance. A third order polynomial was fit to the data in d) demonstrating a good balance between bias and variance. The model accurately captures trends in the data while not overfitting the training dataset.}
	\label{biasvariance}
\end{figure*}

Now that basic methods of machine learning and associated error metrics have been defined we proceed to introduce how machine learning models are fit and validated. The following discussion focuses on finding parameters to fit a machine learning algorithm, how those parameters are validated, and common obstacles that arise in validating the model.

The cost function\footnote{Also sometimes called the loss function or reward function depending on if the objective is to minimize or maximize the value \cite{CostFunction}.} $C({\bf x}; \boldsymbol{\theta})$, is the metric that quantifies the cost of a particular model parameterization. That is, for every input dataset $\mathbf{x}$ there is an associated set of parameters $\boldsymbol{\theta}$ for the machine learning model that best fit $\mathbf{x}$ to their associated outputs $\mathbf{y}$. The training step is concerned with finding the model parameterization that minimizes or maximizes the cost, depending on the application. There are many different choices for cost function and each machine learning algorithm will use its own specific methodology. Perhaps the best known loss function is the squared loss, given in Equation \ref{leastsquares}. 

The loss function is used to minimize a \textit{parameterization} of a machine learning algorithm. For example, a least squares regression algorithm is parameterized by the weighting constants $\beta_i$,
\begin{equation}
	\hat y = \beta_0 + \beta_1 x + \beta_2 x^2 + \ldots + \beta_n x^n.
	\label{beta}
\end{equation}
The model parameters are the weights $\beta_i$ that are fit to the linear regression. The goal of training a machine learning algorithm is to find model parameters that minimize the loss function. If the values of $\beta_i$ are optimally chosen then the value of $\mathbf{X}\beta-\mathbf{Y}$ in Eqn. \ref{leastsquares} should be minimized. The actual method of performing this optimization can take on many forms and is discussed in-depth elsewhere. The scikit-learn package, part of the Python scipy library, provides many methods for optimization of cost functions\cite{ScipyOptimization}. Gradient descent is a common method for performing cost function optimization\cite{GradientDescent2017}. An article by Brochu et al. discusses optimization of cost functions using Bayesian optimization, an important topic in modern statistics \cite{Brochu2010}.

Certain machine learning methods -- such as neural networks, decision trees, and ridge regression -- also have model \textit{hyperparameters}. These parameters define aspects of the model itself, not aspects of a specific parameterization of the relationships between $\mathbf{x}$ and $\mathbf{y}$. For linear regression of a polynomial function to a dataset the weights $\beta_i$ are model parameters and the order of the polynomial is a model hyperparameter. Hyperparameters will be discussed more in-depth later as specific machine learning algorithms are introduced in Section III.

All machine learning models follow a basic training and validation process:
\begin{enumerate}
    \item Divide data into training, test, and validation data: $\{{\bf X}, {\bf y}\} \to \left\{ \{{\bf X}, {\bf y}\}_{\rm train}, \{{\bf X}, {\bf y}\}_{\rm test}, \{{\bf X}, {\bf y}\}_{\rm validate}\}  \right\}$.
    \item Estimate the model parameters, $\hat{\boldsymbol{\theta}}$, using $\{{\bf X}, {\bf y}\}_{\rm train}$ using an appropriate cost function. 
        \item Adjust the model hyperparameters using $\{{\bf X}, {\bf y}\}_{\rm test}$ based on the accuracy of the best fit parameterization of $\boldsymbol{\theta}$.
    \item Validate the best parameterization and check against over- or under-fitting by evaluating the model on the validation set $\{ {\bf X}, {\bf y}\}_{\rm validate}$.
\end{enumerate}
These steps are repeated until the model performance, as measured by the model error estimate, converges.

\begin{figure}
	\includegraphics[width=1\linewidth]{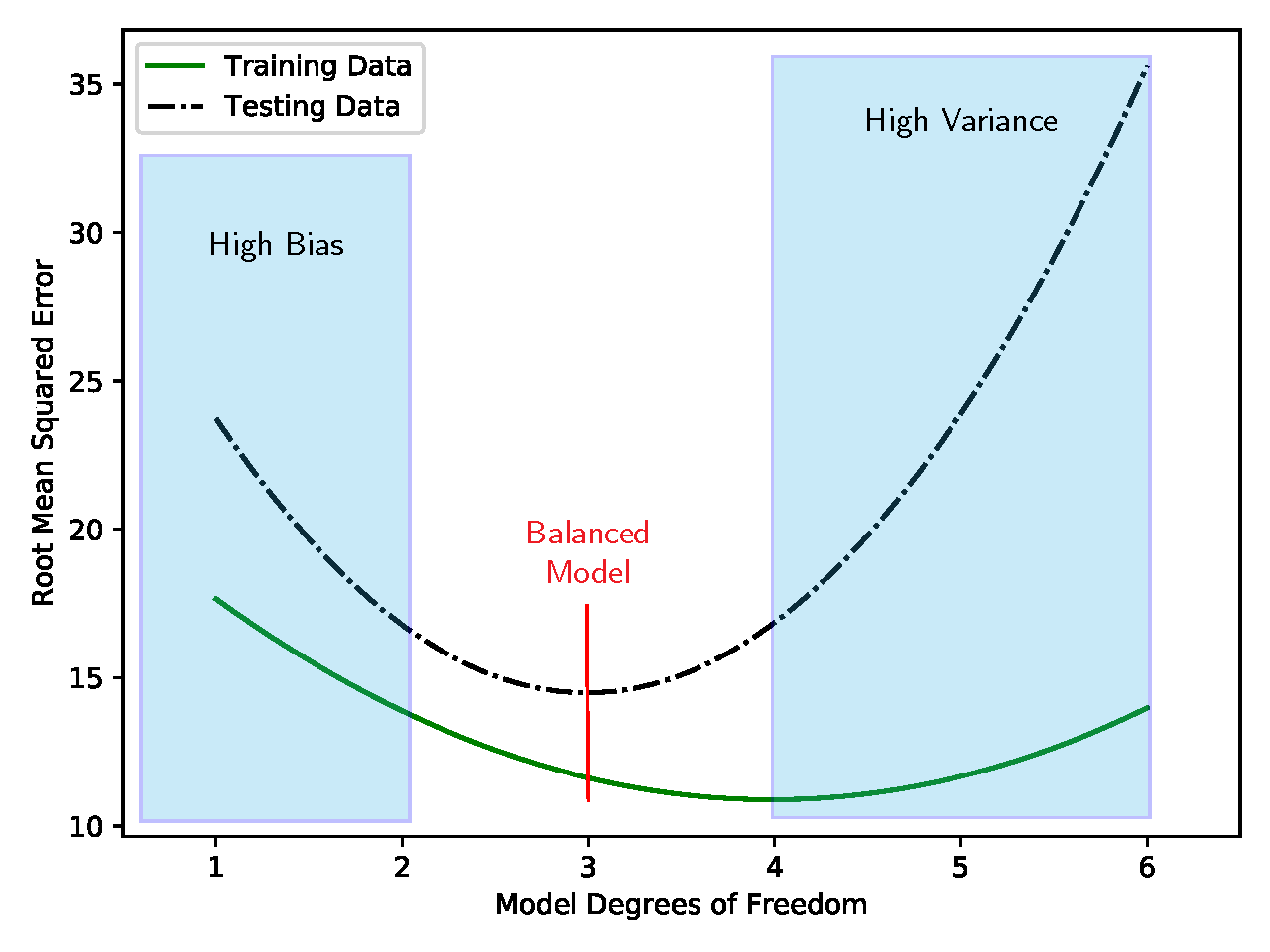}
	\caption{The calculated root mean squared error for six different models fit to the dataset shown in Figure \ref{biasvariance}. The training data continues to decrease as the models become more complex, demonstrating overfitting. However, when the model is evaluated against a test dataset the RMSE increases significantly with more complex models.}
	\label{tvtrmse}
\end{figure}

As the complexity of the model increases -- such as the complexity of a polynomial in a linear regression problem --  so does the tendency of that model to overfit to the training data and generalize poorly to unseen inputs, leading to an increase in the out-of-sample error. This balance between the ability of the model to represent the inherent complexity between the input and output spaces (i.e., reduce the \emph{model bias}) while minimizing the out-of-sample error (i.e., reduce the \emph{model variance}) is the basis for the \emph{bias-variance tradeoff} that is central to all machine learning models. Visual examples of overfitting, underfitting, and proper fitting can be seen in Figure \ref{biasvariance}. The goal in validating a machine learning model is to find a balance between overfitting the training dataset and underfitting the testing dataset, as shown in Figure \ref{tvtrmse}. While the RMSE shows a decrease in the training dataset as model complexity increases, the RMSE of the testing dataset increases significantly.

Overfitting and selection bias can be sussed out through use of \textit{cross validation}. Cross validation is the process of training machine learning models on subsets of the training set and evaluating with the remaining data to see how sensitive the model performance is to the choice of different inputs. Cross validation is often referred to as $k$-fold cross validation because the machine learning model is trained on $k$ different subsets. 

In $k$-fold cross validation the training data set is randomly split into $k$ different groups or \textit{folds}. The machine learning model is trained on $k-1$ of the folds and tested on the $k^{\text{th}}$ fold. For each training and testing set the fit model parameters and associated error should be kept in order to assess how each parametrization of the model performs. 

Randomizing, training, and validating on multiple subsets of the data elucidates the model's ability to perform on new datasets. If a model is suffering from overfitting, or high variance, then it will have very low predictive error on the training set but perform poorly on the testing set. If the model is suffering from high bias then it may demonstrate similar performance metrics between fitting of each $k^{\text{th}}$ set but has high prediction error in general. High bias often results from improper assumptions in the machine learning algorithm or a poor choice of model hyperparameters. Cross validation reveals these behaviors in machine learning models by providing error metrics for models trained on many different subsets. Necessary changes to the model hyperparamters, or even changes in machine learning modeling used, can be discovered from cross-validation.

One specific case of cross validation where $k=n$ is called \textit{leave one out} cross validation. In this method the models are trained on all data points except one, then tested on the remaining data point. Leave one out cross validation is especially useful for assessing the impact on outliers of the model performance. 

Another method of cross validation called leave-one-cluster-out (LOCO) cross validation was introduced by Meredig et al. \cite{Meredig2018} for materials science applications. LOCO CV was introduced to highlight problems in the distribution of data in materials datasets. Often, datasets from materials science are limited around specific clusters of material compositions or properties. An example for AM is that most datasets generated focus around weldable alloys like 300 series steels, superalloys, and titanium alloys. As a result the prediction performance of machine learning algorithms may be biased toward these clusters of materials. LOCO CV uses a nearest-neighbor clustering approach -- akin to the example given in Section \ref{unsupervised} -- to evaluate the impact of clustering of material types on prediction performance.

The above methods are for the validation of individual machine learning models. In many cases it is worthwhile to train several different machine learning models on the same problem and assess the best model. As is shown in Table \ref{ML}, several different machine learning algorithms can often be applied to the same task. Because each algorithm has different assumptions, one type of ML model may perform better on a dataset than others. Thus, it is worthwhile to use tools that can compare the performance of different ML models for the same application.

%
\subsection{Comparison Across Machine Learning Approaches}\label{comparison}
The validation of a single machine learning model can be addressed by the methods presented in Sections \ref{errormetrics} \& \ref{bvar}. Finding the best possible parameterization of an individual model does not guarantee that a researcher has found the best possible solution to their specific problem. It is generally good practice to evaluate several machine learning approaches to a problem and choose the best approach across all algorithms that may be reasonably expected to perform.. Table \ref{ML} shows that many different algorithms can be used for the same types of problems. Different algorithms may have vastly different performance even for the same problem or dataset.

For example, Principal Component Analysis (PCA) and kernel ridge regression (KRR) can both be used as regression tools; PCA relies on the assumption of linearity between inputs and outputs while KRR does not. Often, a researcher might not know the if the relationship being studied is linear or not and therefore should try both options to see which produces a better result.

In general, researchers can follow a few steps to determine which model is best for their additive manufacturing problem:
\begin{itemize}
	\item Evaluate if there are statical correlations in the data of interest 
	\item Pre-process and featurize data for use with a machine learning algorithm
	\item Tune the model parameterization and hyperparameterization through error analysis and cross validation
	\item Compare error metrics across several algorithms and select one algorithm as the best performer
\end{itemize}

\begin{figure}
	\includegraphics[width=1\linewidth]{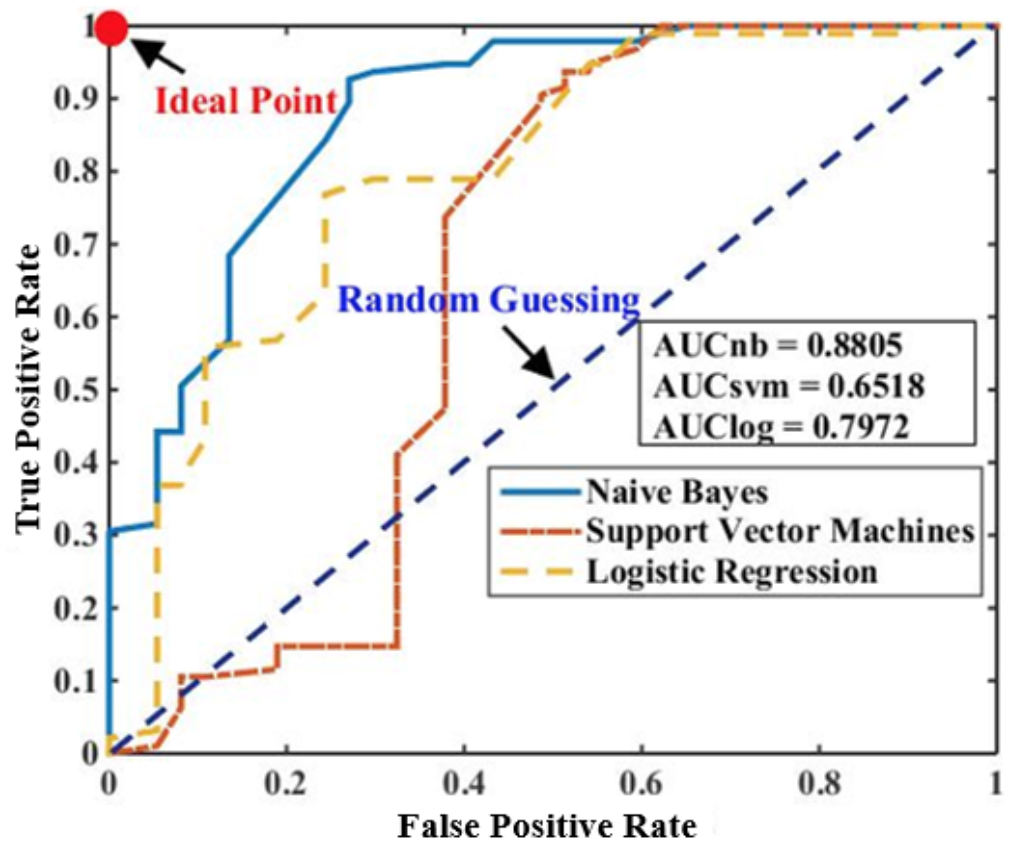}
	\caption{An example receiver operating characteristic curve from the work of Liu et al. \cite{Liu2020}. The goal of the study was to class material properties based on additive manufacturing machine inputs. The classes were regimes of material quality like ``high density" or ``low density." The dataset was built by mining data from literature on additively manufactured metals. The area under the curve (AUC) shows the integrated area under each algorithm's ROC curve; a perfect classifier has AUC$=1$. In the example shown, Na\"ive Bayes significantly outperforms the other two algorithms and thus is the best choice of machine learning approach for this problem.}
	\label{ROC}
\end{figure}

Regression models can be validated against each other using the error metrics in Section \ref{errormetrics}. It is important to use multiple error metrics for comparison because different machine learning algorithms handle outliers and statistical correlations differently. For classification problems, a graph called a receiver operating characteristic (ROC) curve has been developed to compare the classification success of different algorithms. An example ROC curve can be seen in Figure \ref{ROC}. The ROC curve compares the true positive and false positive classification rates for a binary classifier, a group of problems whose solution can take one of two outcomes. To ensure that Type I error (false positive) accurately reflects the performance of the model, the less common outcome should always be taken as the True condition, and the more common outcome as the False condition. (Footnote. Although restricted to binomial classification, the ROC curve may be extended to multinomial classification by recursion. That is, A or not A; and if not A, then B or not B; and if not B, then C or not C; etc. where A, B, C, etc. are all potential outcomes in order of increasing frequency.)More information on ROC curves can be found at Google's developers page \cite{GoogleROC}.

Tools to compare across machine learning algorithms are invaluable and should be considered as a mandatory part of any machine learning approach. It is often the case that evaluating many machine learning algorithms against each other will lead to better overall performance because the best approach can be chosen from many. The ML packages listed in the next section all contain tools for comparing machine learning algorithm performance.

\subsection{Machine Learning Toolboxes}
Most of the machine learning algorithms and approaches discussed in this review are, in some form, free and openly accessible. Many machine learning packages exist across many different programming languages and platforms. Table \ref{data_tools} highlights a variety of computational tools and packages and their relevance to AM synthesis optimization. 

\begin{table*}
        \caption{Commonly-used machine learning, statistical analysis, and computer vision toolboxes. Some toolboxes listed are open source, while some are packaged with commercial software like MATLAB.}
    \renewcommand{\arraystretch}{0.8}
    \setlength{\tabcolsep}{5pt}
    \begin{center}
        \begin{tabular}{p{3cm}p{3.5cm}p{11cm}} \hline
            \toprule
            Language/Platform & Package & Applications  \\ \midrule 
            \hline
            Python & scikit-learn \cite{sklearn} & General data mining toolbox; packages for classification, regression, clustering, dimensionality reduction, model selection, and data pre-processing. \newline \\
            		& tensorflow \cite{tensorflow} & Machine learning toolkit for data mining and data flows; specifically focuses on the use of neural networks and deep learning for model building and problem solving. \newline \\
			& keras \cite{keras} & Deep learning-specific machine learning toolbox; designed for intuitive building of neural network systems. \newline \\
			& OpenCV \cite{opencv} & Algorithm toolbox for machine learning and computer vision; contains wide range of tools for image processing including image pre-processing, template matching, object identification, and convolutional neural networks. \newline \\
			
	   MATLAB & Statistics and Machine Learning Toolbox \cite{matlabml} & Commercial data analysis and machine learning toolbox with a wide range of applications in data analysis including clustering, classification, regression, and dimensionality reduction. \newline \\
	   		&\raggedright Computer Vision Toolbox\cite{matlabcv} & Algorithm toolbox for machine learning and computer vision; contains tools for a wide range of image analysis including pre-processing, object identification, template matching, and convolutional neural networks. \newline \\ 
			
	  C $++$ & OpenCV \cite{opencv} & Algorithm toolbox for machine learning and computer vision; contains wide range of tools for image processing including image pre-processing, object identification, template matching, and convolutional neural networks. \newline \\
	  	& tensorflow \cite{tensorflow} & Machine learning toolkit for data mining and data flows; specifically focuses on the use of neural networks and deep learning for model building and problem solving. \newline \\ 
		
	R & Machine Learning in R (MLR) \cite{mlr} & Infrastructure for incorporating common machine learning functions in R in an easy way; provides robust packages for a wide range of machine learning-based tools including regression, classification, clustering, sampling methods, model optimization and more; has built in parallelization methods. \newline \\  \hline
	
            \bottomrule
        \end{tabular}
        \label{data_tools}
    \end{center}
    
\end{table*}

\section{Current ICME Tools are Well Equipped to Integrate with an ML Framework}

The following section discusses how machine learning approaches can be used in current R\&D efforts in AM. This discussion includes how physics-based analyses, characterizations, and simulation methods may connect with different machine learning algorithms. Overall, the discussion is aimed at conveying how ML can be used to automate the generation of AM PSPP knowledge. Still, this article stops short of providing an exhaustive review of either machine learning algorithms or additive manufacturing. Instead, the intent is to introduce how ML approaches can be connected to AM research. The algorithms that are discussed were chosen because they were previously demonstrated in a materials science and engineering application \textbf{or} because the possible application of an algorithm to AM was clear and immediate. Similarly, the additive manufacturing problems addressed are not all-encompassing; they are merely a few that may be immediately addressable with machine learning approaches. 

\subsection{Experimental Methods and Manufacturing Design}
\subsubsection{Alloy Design and Feedstock Selection}
\begin{figure*}
	\includegraphics[width=1\linewidth]{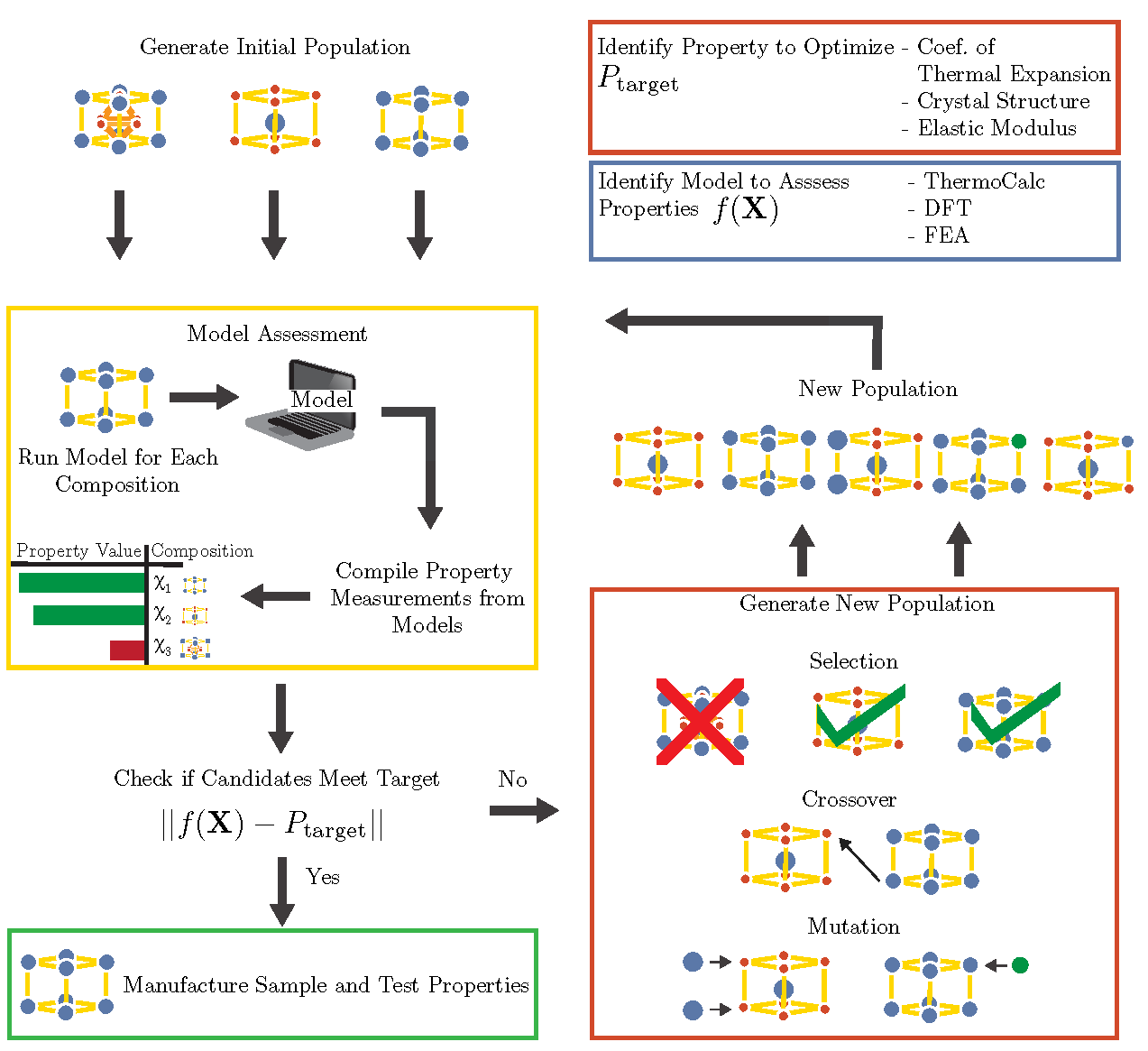}
	\caption{An illustration of the alloy design process using a genetic algorithm. First, a target property $P_\text{target}$ and an evaluation method $f(\mathbf{X})$ for the alloy $\mathbf{X}$ are chosen. The evaluation method is most often a material modeling approach that can predict material properties based on composition. Then, a population of starting compositions are made. The model is run for each composition and an associated material property is measured. The predicted values are compared against the target value. If no material matches the target, then the genetic algorithm begins. The closest-matching compositions are selected to create a child generation. Crossover and mutation occurs for those compositions that were selected. In this way, a new population of compositions are created that are similar to the best-performing compositions from the previous generation. Model assessment and the genetic algorithm are then run again until a composition is found that meets the target property value.}
	\label{fig:GA}
\end{figure*}
Choice of alloy impacts the physics of AM from start to finish, ranging from the interactions of energy sources with material feedstocks to the performances of the final parts. For example: the reflected vs. absorbed intensity of lasers on powder beds is determined by the powder's composition \cite{Boley2016, Trapp2017}; the density of feedstock, both intra- and inter-granular density, plays a role in final part density \cite{Bi2013}; conduction modes in the melt are partially determined by the thermal properties of the alloy \cite{Martin2017}; and different alloys exhibit different solidification kinetics, which can lead to drastically different microstructures after manufacture \cite{Collins2016}. Problems in the additive process can also be linked to composition such as vaporization of constituent elements due to rapid thermal fluxes, impacting the stoichiometry of melt pools and, ultimately, quality \cite{Brice2018}. These can be different for different feedstock types (e.g., wire vs. powder), even for the same alloy choice. Wysocki et al. discuss the differences between different additive manufacturing processes for titanium alloys: electron beam, laser based, powder, wire, etc\cite{Wysocki2017}. Some studies have also investigated the impact of feedstock properties like particle size distribution and morphology on process quality \cite{Slotwinski2014, Strondl2015, Trapp2017}, although the direct impacts have not been fully resolved.

As such, alloys developed for traditional metals manufacturing techniques such as casting, rolling, extrusion, etc. sometimes need to be altered to improve AM processing. In the best cases, alloys developed for AM may outperform traditionally manufactured alloys. For example, unique strengthening mechanisms can result from AM processing \cite{Brice2018, Wang2017, Martin2017, Gallmeyer2019}. Designing alloys for AM -- either altering the chemistries of known alloys or discovering new alloys -- requires considering the implications of the physical properties of alloys with AM processing. An understanding of what trend in a physical property is ``better" or ``worse" for AM processing is still an open area of research. Hence, while information about the physical properties of different alloys has been collated into databases that are compatible with design for AM, models and optimization targets for mining those databases to extract candidate alloys for AM are still being developed and verified. 

Existing databases contain alloy properties ranging from the reflectivity to the mechanical properties. The International Crystal Structure Database (ICSD) contains the crystal structures of millions of compositions. The Linus Pauling files contains a range of material information, from atomic properties like radius and electron valency to crystallographic level information \cite{Villars1998}. More modern databases such as AFLOWLib \cite{Curtarolo2012a} and the Materials Project \cite{Jain2013} allow users to interactively search across different types of alloy information. Searching through large databases of information to find optimal compositions for manufacturing is actually one of the earliest materials informatics problems ever addressed. Methods exist to perform these searches in a fast, automated way. These methods are referred to as data mining, a data-driven materials design approach.

Data mining has been demonstrated to be useful for AM alloy development. Martin et al. used such an approach to modify the chemistry of aluminum alloys to make them process better during LPBF\cite{Martin2017}. The first step in a data-driven design process is to identify which alloy properties are important to the desired application. Laser powder bed fusion of Al alloys had been plagued by sparse nucleation of grains. The result was that large grains formed during AM together with large intergranular stresses, the combination of which resulted in hot-cracking. To overcome this problem, Martin searched for candidate grain inoculant compounds that could form through chemical reactions during LPBF. Searching for grain-refining nanoparticles has improved solidification properties\cite{Neuchterlein2016}. For example, silicon and carbon could react to form SiC particles that would force more homogeneously, densely packed grain nucleation throughout the material. However, if such compounds had lattices that were dissimilar to those of the aluminum alloy, large stresses could form at the interface of the inoculants and the alloy matrix, still leading to cracking. Hence, they searched not only for potential inoculants, but more specifically for inoculants with crystallographic lattice parameters that closely matched those of the base aluminum alloy. Martin's study employed a search algorithm to search through 4,500 different possible nucleants and identify those with the closest-matching parameters. Ultimately, hydrogen-stabilized Zr was found to be the best candidate.

The same database mining process employed by Martin -- identify the target properties, then search for the closest match -- can be extended to many AM problems as well. Database mining was first introduced in materials science to predict stable compositions, or estimate material properties from composition. Database mining has been successfully implemented to predict stable crystal structures \cite{Franceschetti1999, Fischer2006, Oganov2006} and predict material properties as a function of composition \cite{Ikeda1997, Gopakumar2018, Wu2018, Kirklin2013, Setyawan2011}. Some specially designed search algorithms have also been designed for improved speed in automated searches \cite{Wolf2000}. Successes have been found in designing Heusler compounds using high throughput search methods \cite{Roy2012}. Several reviews exist detailing early high-throughput searches for compositions with ideal properties\cite{Gilmer1998, Koinuma2004}. The same search algorithms employed in these studies can be extended to AM cases.

A limitation of database mining is that searches are limited to previously measured and/or calculated properties. Generally, information about the vast space of \textit{all possible} materials is unknown. Traditional materials science and engineering approaches would turn to explicitly calculating or measuring the unknown points of interest, one at a time. Searching through compositions may be accessible for manufacturing processes like thin-film deposition where the composition can be adjusted continuously and with several species at once using well established methods. A combinatorial study of compositional changes for AM feedstock is hindered by the difficulty and expense of producing feedstock.

For example, consider the cost of combinatorially alloying Ti with alloying elements $\{\text{Al}, \text{V}, \text{Zr}, \text{Cr}, \text{Hf}\}$ and then testing printability. Explicitly creating all possible combinations of $\{\text{Ti},\text{Al}, \text{V}, \text{Zr}, \text{Cr}, \text{Hf}\}$ is feasible if using a coarse set of level choices for additions of alloying elements, but undesirable. There are 15,503 alloy combinations if alloying in steps of $1$ wt. \% up to $15$\% total alloying elements from the choices above.

However, using machine learning methods, the process of combinatorial exploration to find an optimal composition can be achieved without explicitly modeling each combination. For example, \textit{genetic algorithms} (GA) can be used to augment many physics-based models. Genetic algorithms have been one of the most-used data driven approaches in materials science over the past few decades \cite{Morris1996, Ho1998, Wolf2000, Johannesson2002, Stucke2003, Hart2005, Oganov2006}. The principle of genetic algorithms is to evaluate the \textit{fitness} of a population of candidate alloys against a \textit{fitness function}. The fitness function $f(\cdot)$ is a method of evaluating how well a candidate alloy meets a criteria. Often in materials science the fitness function is evaluated by running models that can measure a material property based on composition. Examples include identifying stable crystal structure of a composition using DFT\cite{Franceschetti1999, Oganov2006} and evaluating thermomechanical properties of an alloy using ThermoCalc \cite{Xu2008}. Some additive-specific models include the model of Tan, which predicts dendrite arm spacing from composition \cite{Tan2011}. The calculation of thermodynamic properties relevant to AM -- such as vaporization temperature, coefficient of thermal expansion, solidus and liquidus temperatures -- using the CALculation of PHAse Diagrams (CALPHAD) method \cite{Andersson2002} can also be a fitness function.  For the sake of alloy design a model must be able to predict a materials properties based on composition. In reality, however, models must also consider additional physics related to the composition, such as crystal structure, thermodynamic properties, interatomic potentials, and more.

In using a GA for alloy design, a desired target property value must be identified. This value $P_\text{target}$ is then formulated as a function of composition and process variables. Additionally, a method of measuring the property value as a function of composition and process variables $\mathbf{X}$ is needed; the models proposed previously (ThermoCalc, DFT, etc.) can serve as the evaluation step $f(\mathbf{X})$. The goal is to find a material whose measured property closest matches the desired target property, or

\begin{equation}
	\text{min} || f(\mathbf{X}) - P_\text{target} ||.
	\label{GAopt}
\end{equation}
As a thought experiment, consider various amounts of $\{\text{Al}, \text{V}, \text{Zr}, \text{Cr}, \text{Hf}\}$ alloyed into Ti. These are the \textit{genes} of the genetic algorithm. This is similar to a study completed by Li et al\cite{Li2017}. Once a fitness function has been identified, the next step in a genetic algorithm is to represent candidate alloys as a \textit{chromosome}. 

We can represent a chromosome as \\

\begin{table}[h!]
\begin{tabular}{cccccc}
	$\mathbf{X}$ & $=$ & [$\chi_1$, & $\chi_2$, & $\ldots$, & $\chi_n$] \\
\end{tabular}
\end{table}
\noindent 
where $\chi_1$ is the species and weight percent of the first element (titanium, in this example), $\chi_2$ is the species and weight percent of the second element, up to $n$ elements. For example, Ti-6Al-4V would be represented as \\

\begin{table}[h!]
\begin{tabular}{ccc}
	 [0.9 Ti,  & 0.06 Al, & 0.04 V ] \\
\end{tabular}
\end{table}
\noindent
The goal is to find the alloy with optimal dendrite arm spacing. First, a population of candidate chromosomes needs to be generated, either randomly or by design. Two examples from a starting population may be \\

\begin{table}[h!]
\begin{tabular}{ccccc}
	Alloy 1 & $=$ & [0.9 Ti, & 0.05 Al, & 0.05 V ] \\
	Alloy 2 & $=$ & [0.9 Ti, & 0.1 Zr] & \\
\end{tabular}
\end{table}
\noindent
The chromosomes produced from this initial population will serve as inputs to the fitness function. 

Genetic algorithms select chromosomes out of the current population -- called the parent generation --  to proceed to another generation of model assessment -- called the child generation. Selection consists of keeping the best performing compositions, say the top $10\%$, and discarding the rest, as determined by Eqn. \ref{GAopt}. Genetic algorithms find optimal locations in the design space by relying on the similarity hypothesis. If one alloy is in the top $10\%$ of chromosomes then it is possible that a similar alloy will also be high performing -- it may even perform better. Once selection is done, the next step is to search the space near the best performing alloys from the parent generation.

Genetic algorithms generate similar compositions from those selected in the parent generation by making alterations to genes. One operation is \textit{mutation}, whereby genes are changed. For example, we could mutate alloy 1 by changing the composition:\\

\begin{table}[h!]
\begin{center}
\begin{tabular}{c|ccccc}
	\textbf{Parent Generation:} & Alloy 1 & $=$ & [0.9 Ti, & {\color{red}0.05} Al, & {\color{red}0.05} V ] \\ \hline
	\textbf{Child Generation:} & Alloy 1 & $=$ & [0.9 Ti, & {\color{blue}0.02} Al, & {\color{blue}0.08} V  ]  \\ 
\end{tabular}
\end{center}
\end{table}
\noindent where in the child generation the amount of V was increased, while the amount of Al was decreased. Another operation that may be performed is \textit{crossover} where genes are added or interchanged. For example, one crossover operation may look like

\begin{table}[h!]
\begin{center}
\begin{tabular}{c|ccccc}
	\textbf{Parent Generation:} & Alloy 1 & $=$ & [0.9 Ti, & 0.05 Al, & 0.05 {\color{red} V} ]  \\
						 & Alloy 2 & $=$ & [0.9 Ti, & 0.1 {\color{blue} Zr}] &              \\ \hline					 
	 \textbf{Child Generation:} & Alloy 1 & $=$ & [0.9 Ti, & 0.05 Al, & 0.05 {\color{blue} Zr} ]  \\
						& Alloy 2 & $=$ & [0.9 Ti, & 0.1 {\color{red} V}] &              \\ 
\end{tabular}
\end{center}
\end{table}
\noindent where in the second generation V and Zr have been interchanged.

Selection, mutation, and crossover followed by model assessment and further selection, mutation, and crossover continues until the design criteria is met. A schematic of the GA process can be seen in Figure \ref{fig:GA}. 

Genetic algorithms have been applied to alloy design for low and high temperature structural materials \cite{Ikeda1997, Kulkarni2004}, ultra high strength steels \cite{Xu2008}, specific electronic band gaps \cite{Dudiy2006}, minimum defect structures \cite{Anijdan2006}, exploring stable ternary or higher alloys alloys \cite{Hautier2010, Johannesson2002}, and more. Chakraborti et al. wrote a review on the application of GA's to alloy design through the early 2000s\cite{Chakraborti2004}.

In addition to genetic algorithms, other machine learning algorithms have also been applied to classify and optimize alloy compositions. Anijdan used a combined genetic algorithm--neural network method to find Al-Si compositions of minimum porosity \cite{Anijdan2006}. Liu et al. applied partial least squares to data mining of structure-property relationships across compositions \cite{Liu2006}. Decision trees, which are discussed in the next section, have been implemented for a number of different alloy optimizations, such as predicting ferromagnetism \cite{Landrum2003} and the stability of Heusler compounds \cite{Oliynyk2016}. In the search for new alloys, a wide range of machine learning algorithms can be implemented to guide the entire experimental design process so that an optimized property is found as quickly as possible. In the next section, we focus on using ML in design of experiments.
\subsubsection{Design of Experiments}
Design of Experiments (DOX) is the design of task(s) aimed at performing parametric analysis \cite{Dox2014}. Parametric analysis, broadly defined, is a method of mapping independent variables to corresponding dependent parameters. In materials science and engineering, process-property relationships are typically assessed using parametric analysis. Machine learning can reduce the number of experiments (i.e., tasks) needed to perform parametric analyses sufficient to characterize process-property relationships. Approaches such as \textit{sequential learning} model relationships in parametric studies to discover regions of the parameter space that produce the most information about process-property relationships. 

 In additive manufacturing research, process parameters such as laser energy, speed, build direction, composition, and layer height are varied to study their impact on material properties. Examples include relating build geometry to microstructure or surface roughness \cite{Antonysamy2013, Strano2013}, temperature history to microstructure \cite{Bontha2009, Nie2014}, substrate temperature to residual stress development \cite{Chen2016, Brice2018}, or even entire manufacturing processes to microstructure \cite{Baufeld2011}. Other commonly performed parametric analyses in AM relate heat source parameters to part temperature history \cite{Bontha2006, Li2014}, microstructure \cite{Cherry2015, Jia2014}, mechanical properties \cite{Delgado2012, Khorasani2018}, and residual stresses \cite{Wu2014, Denlinger2015}.

\textit{Information} in AM research is any observation of process-structure-property relationships. For example, observing that a set of laser parameters results in an equiaxed microstructure can be considered information because the researcher has gained an idea of the structure to expect from set processing conditions. Therefore, \textit{information gain} is any experiment that reveals a previously unobserved process-structure-property relationship. Rigorous mathematical definitions of information and information gain have been defined, typically referencing back to Shannon's original formulation of information theory \cite{Shannon1948}.

Both engineering and scientific investigations of AM utilize parametric analysis. In science, tasks designed for information gain are performed until parametric analysis results in a theory or model for a process-structure-property relationship. In engineering, tasks designed for information gain are performed until an optimality criterion is met, such as maximum strength or minimum porosity. Both disciplines vary independent parameters and measure dependent responses until enough information about the underlying phenomenon is known to complete the parametric analysis with some predetermined level of certainty, variance, and/or precision. 

Traditional DOX approaches maximize information gain from performing tasks by subdividing the design space \textit{a priori} to maximize the likelihood of information gain from task to task. In these approaches, all pre-determined tasks are performed before parametric analysis is attempted. In machine learning DOX approaches, parametric analysis is performed after each individual task, and the next task to perform is determined based upon a statistical metric of the parametric analysis - as such, the likelihood of information gain incrementally improves as each task is carried out, and usually only a fraction (20 - 60\%) of the number of tasks need to be performed to reach the established success criterion for the parametric analysis relative to the traditional DOX approaches \cite{Wigley2016, Ling2017a}.

For ML-based DOX, the first step is still to identify process-structure-property parameters of interest and to classify them as either inputs or outputs relative to the desired relationship that is to be determined, as is done in traditional DOX. As more parameters are added, the size of the design space grows. Once the scope of the design space has been defined, the next step is to generate an initial dataset (i.e., initial information). The first tasks can be designed with traditional DOX methods -- often, an approach as simple as selecting an initial uniform sample from the design space. In addition to generating an initial dataset, a \textit{response function} must be defined to interpret the relationship between the inputs and outputs. One example is a regression model of the process parameters (inputs) and material properties (outputs). A \textit{random forest} algorithm trains many regression algorithms, each on a subset of the experimental data. 

Random forest algorithms are ensembles of a type of simple regression algorithm called a classification and regression tree or a decision trees. Decision trees can be used for both classification and regression. Consider a design space that an engineer wishes to explore represented as a matrix, such as \newline

\begin{center}
\begin{tabular}{c|c|c|c} 
	Feature 1 & Feature 2 & Feature 3 & Property 1 \\ \hline
	$x_{1,1}$ & $x_{1,2}$ & $x_{1,3}$ & $y_{1}$ \\
	$\vdots$ & $\vdots$ & $\vdots$ & $\vdots$ \\
	$x_{n,1}$ & $x_{n,2}$ & $x_{n,3}$ & $y_{n}$ \\
\end{tabular}
\end{center}
where $x_{1,1}$ is the first parameter setting for feature 1, $x_{1,2}$ is the first parameter setting for feature 2, and $y_1$ is the first property measurement for the associated position in the design space, out of $n$ total measurements. This design space could be represented as a matrix by
\begin{equation}
	\mathbf{B} = \begin{bmatrix}
		x_{1,1} & x_{1,2} & x_{1,3} & y_{1} \\
		\vdots & \vdots & \vdots & \vdots \\
		x_{n,1} & x_{n,2} & x_{n,3} & y_{n} \\
	\label{Bmatrix}
	\end{bmatrix}
\end{equation}
The rows of $\mathbf{B}$ represent different observations in the design space and the columns of $\mathbf{B}$ are different parameters or properties. The goal of parametric analysis is to map different values of $x$ to a property $y$. 

Decision trees begin by taking a samples from the design space -- rows in $\mathbf{B}$ -- and computing a split in one of the features (columns) that best classifies the data point. Consider a set of three experiments that has feature-property $\left(x,y\right)$ pairings of $\left(0.1, 0\right)$, $\left(0.2,0\right)$ and $\left(0.3,1\right)$. The decision tree computes every possible partitioning of $x$ and computes a misclassification error called the Gini impurity, defined as 
\begin{equation}
	I_G(x) = \sum_i^J p_i\left(1-p_i\right)
\label{gini}
\end{equation}
where $p_i$ is the percentage of samples classified into class $i$ for each split out of $J$ classes. For our fictional example, $J = 2$.  Consider a split along the value $x = 0.1$ where values less than or equal to $0.1$ are predicted to have $y=0$ and values of $x$ greater than $0.1$ are predicted to have $y=1$. The Gini impurity can be calculated for each side of the split. For the case of $x\leq0.1$, all the samples provided (only one sample, in this case) have an associated $y$ value of $0$. Therefore, the Gini impurity would be 
\begin{equation}
	\begin{split}
	I_G\left(x \leq 0.1\right) & = \frac{1}{1}\left(\frac{1}{1} - 1\right) + \frac{0}{1}\left(\frac{0}{1} - 1\right) \\
		& = 0.
	\end{split}
\end{equation}
The value $0$ is the lower bound for the Gini impurity, thus this split produces perfect classification for values sorted into $x \leq 0.1$. However, the Gini impurity for the remaining values becomes 
\begin{equation}
	\begin{split}
		I_G\left(x > 0.1\right) & = \frac{1}{2}\left(\frac{1}{2} - 1\right) + \frac{1}{2}\left(\frac{1}{2} - 1\right) \\
		& = 0.5.
	\end{split}
\end{equation}
This higher value of the Gini impurity indicates that splitting feature $x$ along the value $0.1$ produces an imperfect classification. If the split was chosen along $0.2$ instead, the Gini impurity for both sides would be $0$, a perfect classification. The Gini impurity can be extended to an arbitrary number of classes, allowing decisions trees to behave as regression algorithms as well as classification tools.

Decision trees compute every possible partition for each feature in the dataset such that the misclassification error, as defined by the Gini impurity, is minimized. However, decision trees are highly prone to overfitting. Random forests overcome this overfitting problem by training many different decision trees, each on a subset of the total dataset. A random sampling, with replacement, of design space coordinates (rows of $\mathbf{B}$) are chosen, known as bootstrap aggregating, or bagging, and a decision tree is trained. Alternatively, or in addition to bagging, jackknifing selects a subset of features (columns of B) to prevent overfitting to specific features..

Training many different decision trees in this way allows a user to calculate uncertainty metrics for each prediction. The method of calculating uncertainty depends on how the random forest is being applied \cite{Ling2017a}. Once the random forest has been trained on the initial dataset, new points in the design space are given to the algorithm and the expected result is predicted. 

The predictions made for new points in the design space can be characterized by several different response functions. A study by Ling et al. employed three response functions: the maximum likelihood of improvement (MLI), maximum expected improvement (MEI), and maximum uncertainty. Each response function has its own benefits. The MEI selects the best experiment for maximizing (or minimizing) a target value. The MU, as the name implies, selects the experiment with the highest uncertainty in predicted result. The MLI chooses the experiment most likely to have a higher (or lower) target value compared to the best previously observed value.

Often, parametric analysis is concerned with either exploring relationships in the design space or optimizing on a property (either minimizing or maximizing the property). The random forest can be trained on $m$ many subsets of the $n$ rows of $\mathbf{B}$. Then, new points in the design space are chosen and their associated property $y_{n+1}$ is predicted. If the goal is to maximize a property, then the next experiment to run can be chosen by the MEI or MLI. If the goal is to explore the design space then it is useful to choose the design space coordinate based on the MU.

\begin{figure}
	\begin{subfigure}{0.5\textwidth}
		\includegraphics[width=1\linewidth]{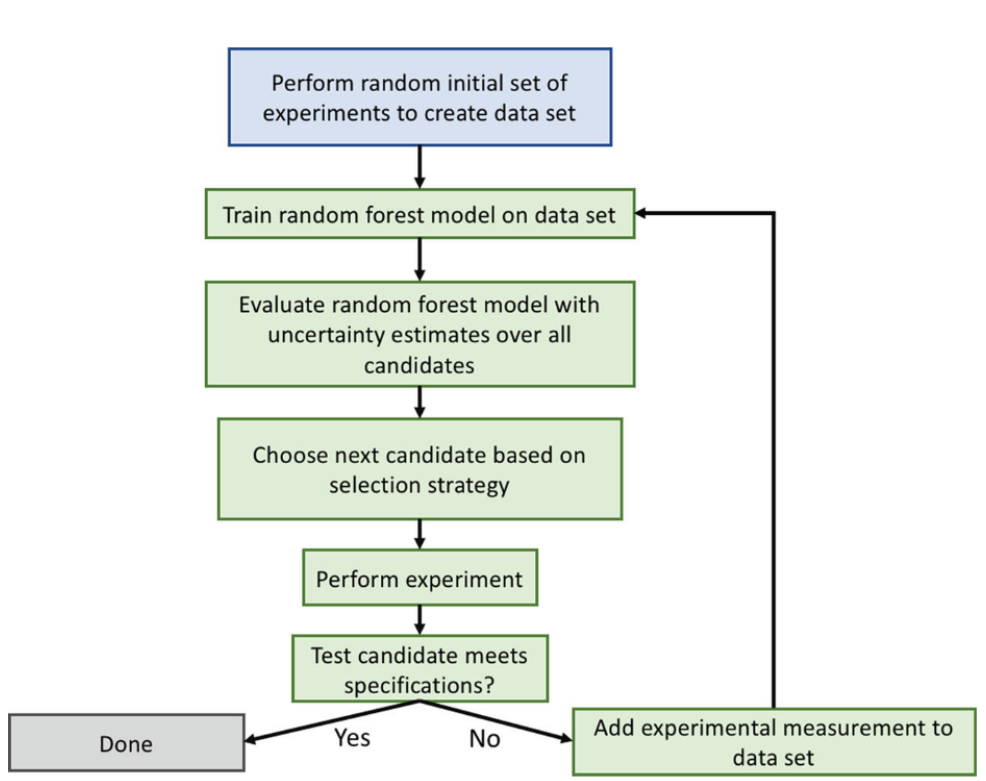}
		\caption{}
		\label{LingAlgorithm}
	\end{subfigure}
	\begin{subfigure}{0.5\textwidth}
		\includegraphics[width=1\linewidth]{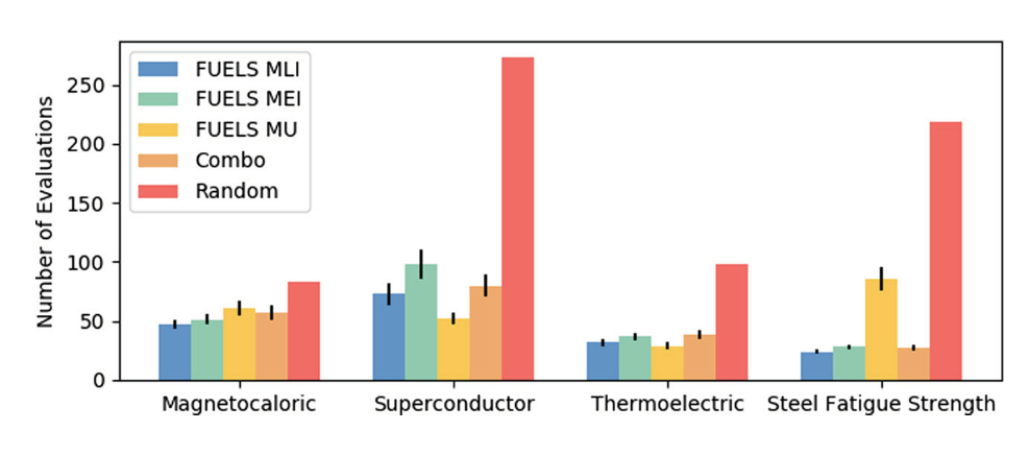}
		\caption{}
		\label{RandomForestOpt}
	\end{subfigure}
	\caption{Application of a random forest algorithm to find optimal material candidates for four different datasets: magnetocaloric materials, superconducting materials, thermoelectrics, and steels. A random forest algorithm was used with four different response functions: maximum likelihood of improvement (MLI), maximum expected improvement (MEI), maximum uncertainty (MU), and the COMBO Bayesian optimization approach \cite{Ling2017, Ueno2016}. The algorithm in \ref{LingAlgorithm} was used to speed up the experimental design process. In every case, the optimal material for the application in the dataset was found more quickly through sequential learning than through random guessing. The random forest approach was compared against COMBO, another sequential learning tool. The figure in \ref{RandomForestOpt} demonstrates how much more quickly the random forest algorithm was able to find an optimized state than random sampling of experiments to perform.}
	\label{RFopt}
\end{figure}

Ling et al. trained a random forest to maximize the fatigue life of steel as a function of composition (among other test cases presented in the article) \cite{Ling2017}. The features used in Ling's study included composition as a function of nine different alloying elements $\left(\text{C, Si, Mn, P, S, Ni, Cr, Cu, Mo}\right)$ as well as thirteen different processing steps such as heat treatment temperature. The total dataset used had 437 tests of steel fatigue as a function of the features. The random forest algorithm was used to choose experiments to run balancing maximum predicted fatigue life with uncertainty in the prediction. Ling's random forest approach found the composition and processing combination with the best fatigue life in fewer than 50 experiments out of the 437 possible options when using the MLI. The sequential learning workflow used by Ling, as well as the performance of differently trained random forest algorithms and response functions is shown in Fig. \ref{RFopt}.

Random forests have been applied successfully to a range of applications in materials science. They have been used to discover new thermoelectric materials~\cite{Gaultois2016}. They have also been used to model material properties such as thermal conductivity in half-Heusler semiconductors ~\cite{Carrete2014} and to break down fields for dielectrics~\cite{Kim2016}.  A review article detailing many optimization algorithms for design of experiments can be found in Shan et al\cite{Shan2010}. Adoption of machine-learning assisted design of experiments algorithms can rapidly increase the rate at which the relationship between AM process parameters and material properties are understood.

\subsubsection{Topology Optimization and Generative Design} \label{sec:topology optimization}

\begin{figure}
	\includegraphics[width=1\linewidth]{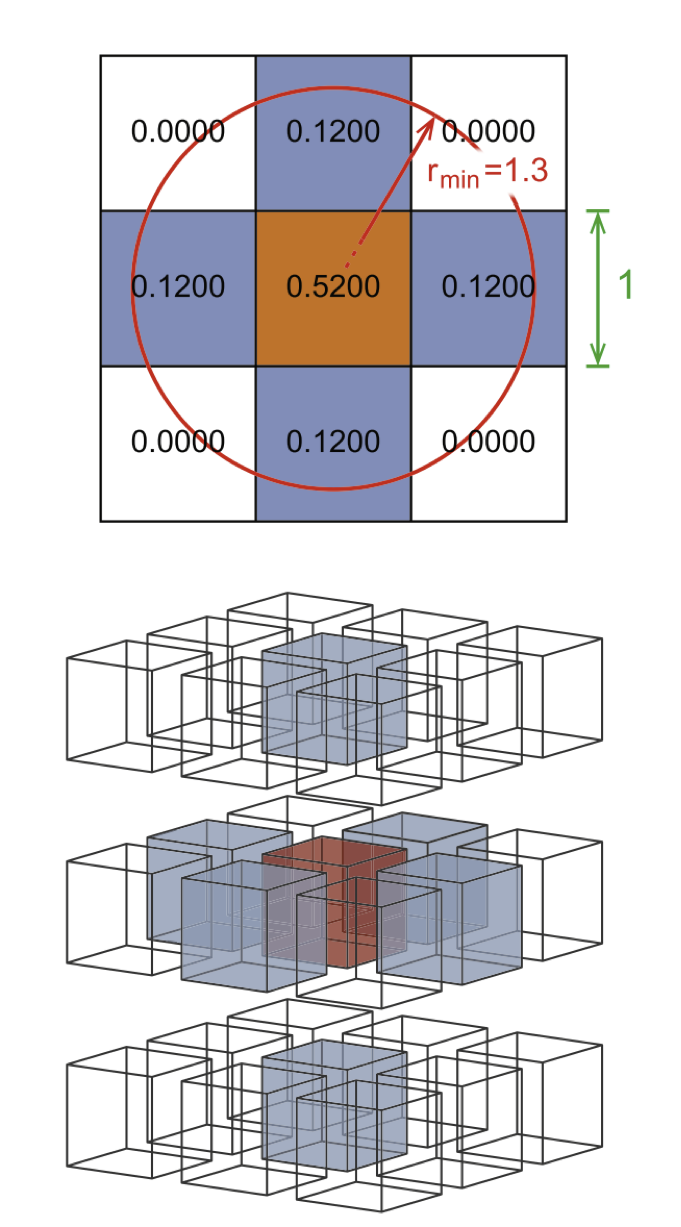}
	\caption{Examples of filters that are applied to CAD meshes to change the geometry of the part \cite{Zegard2016}. The filters can be applied to remove material for weight reduction or add material to prevent part warpage during manufacturing.}
	\label{topfilter}
\end{figure}

Alloy design and process design are based upon process-structure-property relationships of materials, independent of part geometries. These optimizations reduce manufacturing costs and times and help attain targeted properties. Analogous optimization approaches can be applied to design the geometry-material-performance relationships of AM parts. Such approaches are called \textit{topology optimization} methods. For structural materials and their parts, a common goal is to optimize the geometry to maximize the load bearing capacity, stiffness, or lifetime while minimizing the mass of the part. The ability to manufacture the unique, complex geometries determined by topology optimization algorithms for (nearly) the same cost as simple geometries designed for subtractive manufacturing processes is one of the greatest promises and appeals of additive manufacturing.  One of the frontiers in research driven by AM processing, in which materials and part topologies are simultaneously manufactured, is to integrate alloy processing optimization with topology optimization to create concurrent optimization methods. Thus, part performance  becomes integral to the material manufacturing optimization process in AM. Hence, we proceed to introduce topology optimization to the materials researcher while also discussing potential uses for machine learning to advance topology optimization.

Topology optimization can be applied for several optimization objectives, including compliance minimization, stress constraint or natural frequency maximization. Manufacturing constraints such as overhangs and support structures found in AM have also been added to the optimization process and improve the applicability of the result \cite{Sigmund2013,Gaynor2016,Langelaar2016, Langelaar2017,Liu2018}. Support structure optimization that minimizes the amount of material used in supports has also been researched \cite{Huang2009, Vanek2014, Dumas2014}. Other additive specific algorithms have been designed for optimizing density of parts \cite{Zegard2016}. By determining the ideal material layout, the final design can maximize performance for a given weight, minimize weight for an objective function, or reduce manufacturing costs by reducing the material used. However, TO is a local optimization method. Global design optimization usually requires statistical analysis of many TO simulations. 

ML can help reduce the computational time necessary for a TO analysis. This approach allows for faster convergence to a TO result, as well as produces multiple designs efficiently for a researcher to better explore the possible design options. The process of producing many outputs for a given set of conditions is known as generative design.

One approach is to change the topology of a region of a part by applying local filters to CAD models; mathematically, these filters are the same ones shown in Eqn. \ref{filter}.  A physical representation of a filter matrix in 2D and 3D can be seen in Figure \ref{topfilter}. Topology optimization proceeds by generating a CAD model of an AM part and modeling its performance, such as testing performance under mechanical load through an FEA simulation. Filters are applied to the CAD mesh that selectively remove material from the part. Then, the mechanical performance of the new part is modeled, followed by further material removal. This process proceeds until either a minimum weight/volume condition is met or the mechanical performance of the part is degraded.

\begin{figure*}
	\centering
	\begin{subfigure}[t]{1\textwidth}
		\includegraphics[width=6in]{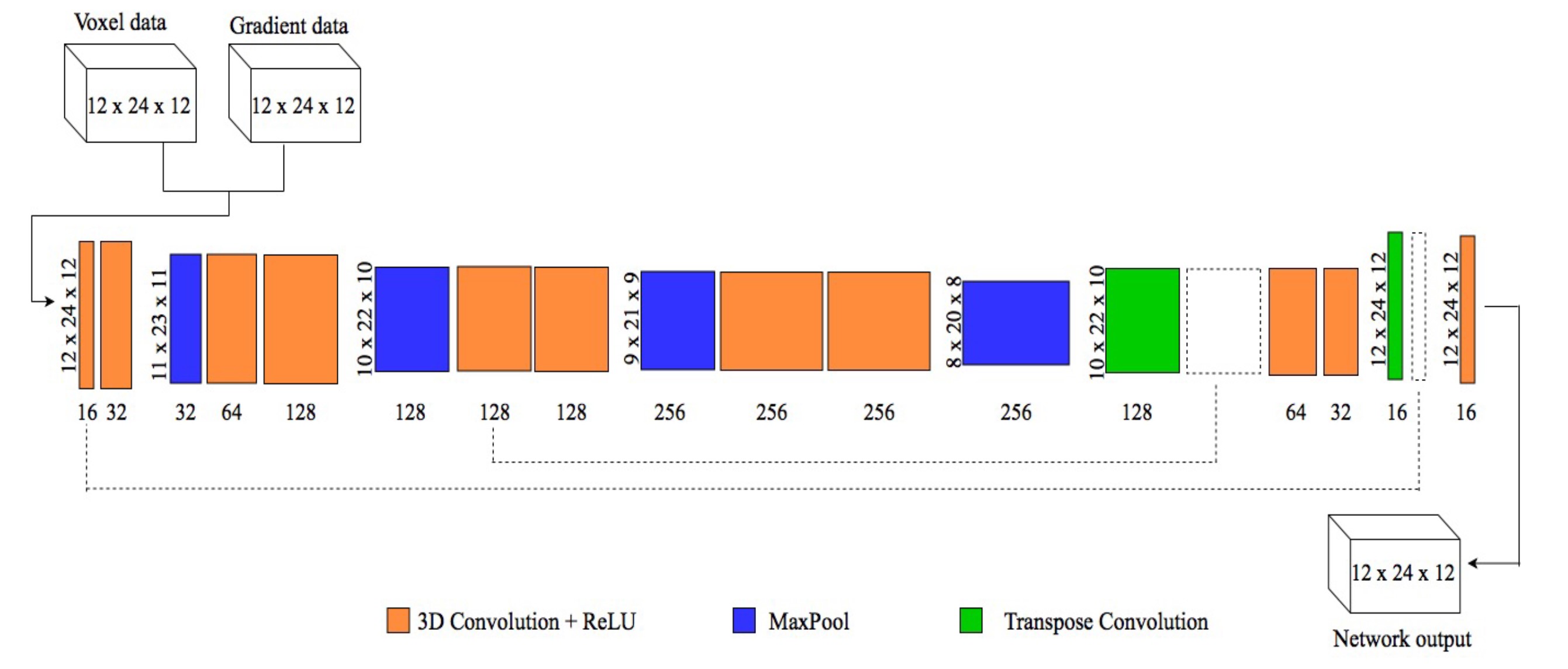}
		\caption{}
		\label{topopt1}
	\end{subfigure}
	~
	\centering
	\begin{subfigure}[t]{1\textwidth}
		\includegraphics[width=6in]{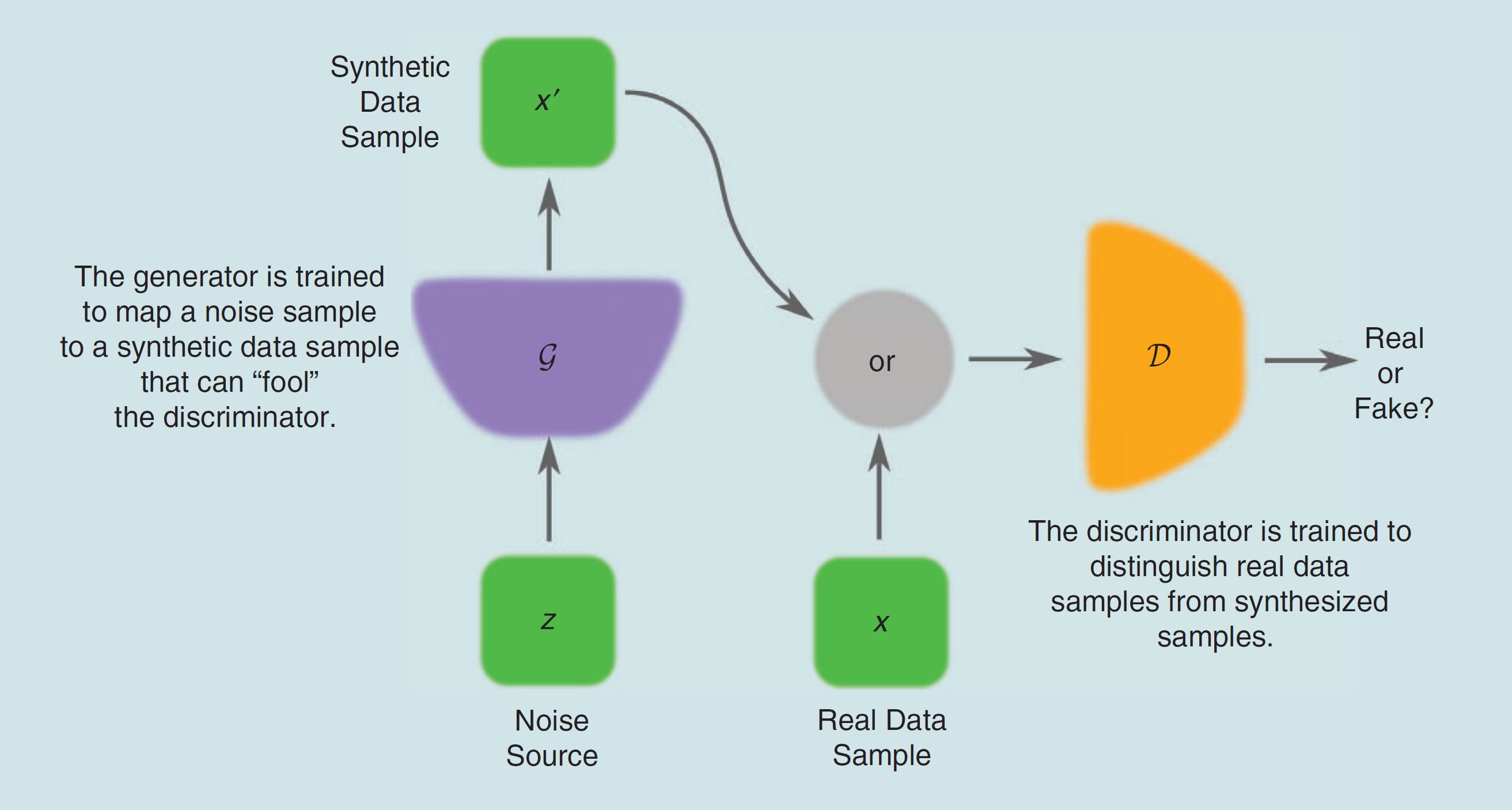}
		\caption{}
		\label{topopt2}
	\end{subfigure}
	~
	\centering
	\begin{subfigure}[t]{1\textwidth}
		\includegraphics[width=6in]{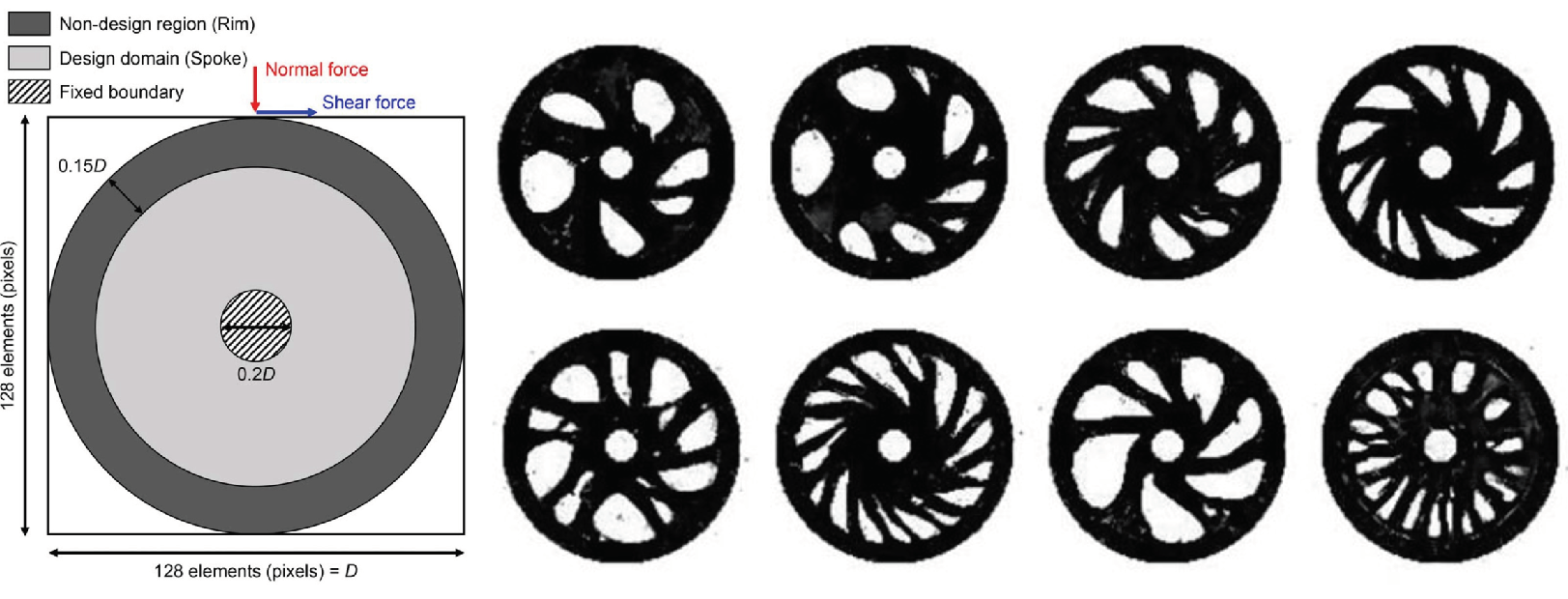}
		\caption{}
		\label{topopt3}
	\end{subfigure}
	\caption{The architectures for convolutional neural networks (CNN) and generative adversarial networks (GAN) are visually described. \ref{topopt1} The network architecture of the CNN used in Banga et al. with the voxel and gradient input data \cite{Banga2018}. The dimensions are described as Height x Length x Width with channels below their respective layer. \ref{topopt2} The models and data used for training a generator and a discriminator in a GAN \cite{Creswell2018}. \ref{topopt3} Examples of Generated designs produced from GAN network in Oh et al. for the design problem shown \cite{Oh2019}. Image in \ref{topopt1} taken from \cite{Banga2018}; image in \ref{topopt2} taken from \cite{Creswell2018}; image in \ref{topopt3} taken from \cite{Oh2019}.}
	\label{topopt}
\end{figure*}

These filters, filters that identify which material to remove, can be learned by a type of machine learning algorithm called a convolutional neural network (CNN). Convolutional neural networks have been found to be well-suited for data containing multiple arrays, especially for image recognition tasks \cite{Lecun2015}. The input is separated into different channels, such as RGB for three channels of a color image input, and manipulated through different stages of the network, called layers. Commonly used layers in these networks include convolutional, pooling, and fully connected. Convolutional layers are divided into varying feature maps that abstract the input to smaller, localized regions for analysis. Pooling layers clusters the outputs from the previous layer and outputs either the maximum or average value from the cluster, reducing the dimensionality of the problem. Fully connected layers connect the outputs from the previous layer with the inputs of the next.

Using a convolutional neural network, Cang et al. and Banga et al. present similar approaches to produce “one-shot” tools for two- and three-dimensional TO, respectively \cite{Cang2019,Banga2018}. One-shot tools produce an optimized structure directly from a starting topology, as opposed to iterative tools that require multiple passes of the algorithm to reach an optimized state. The inputs of the CNN were aspects of the initial part geometry and expected loading conditions. Features given to the model included the force experienced by the part, fixed boundary conditions, minimum mass and density values, and the locations of mass in the part. Their training and validation databases a dataset of optimized topologies generated via traditional TO. The goal of the CNN was to predict an optimized geometry for a starting structure in one pass through the model using knowledge of the loading conditions and part geometry. The results from both works show similar accuracy between the “one-shot” result and the ground truth found from traditional non-ML methods. Such methods greatly reduce the computational time, allowing for greater design exploration before finalizing the result. The architecture for the CNN used in Banga et al. can be found in Figure \ref{topopt1}.

An extension of the convolutional neural network is the generative adversarial network (GAN). The methodology for GANs involves two neural networks in competition with each other: a generator and a discriminator. The generator attempts to create data similar to that of an existing database. The discriminator has access to the database and discerns which data samples are from the database and which are from the generator \cite{Creswell2018}. The goal of the generator is to generate data that the discriminator cannot accurately classify into database or generated datasets. As the discriminator improves at discriminating between artificially generated data and user-provided data, the generator learns to produce better and better artificially generated data. The ultimate goal is a generator network that learns to produce high-quality optimizations of an input topology. A flow diagram of a GAN from Crewswell et al. can be found in Figure \ref{topopt2}. Further information about the applications of GANs, including image synthesis and superresolution, can be found in Crewswell et al. \cite{Creswell2018}.

Yu et al. uses a combined CNN-GAN to perform a superresolution for TO, upscaling a coarse mesh result to a higher resolution without the added computational time to directly compute the high resolution result \cite{Yu2019}. First, a CNN was trained to predict low-resolution optimal geometries based on provided boundary conditions. The CNN used information such as minimum mass fraction, location of applied load, and fixed boundary conditions to predict the best topology at a low resolution. Then, a GAN was trained with random sampling of low and high resolution TO results as inputs and ground truth database, respectively. The GAN was trained to generate high resolution topologies from low resolution inputs. The low resolution output of the CNN was given to the generator; the generator then produced a high resolution topology from the given low resolution input \cite{Yu2019}. The results showed very high agreement between the generated structures and a set of training structures generated using an open source code \cite{Andreassen2011}. The result from this combined network was within 3\% of the expected ground truth pixel values and produced it in 0.06\% of the time compared with traditional TO \cite{Yu2019}.

For generative design, genetic algorithms and GANs are well suited as both architectures are designed to produce multiple optimal designs. Lohan et al. and Zimmerman et al. use the genetic algorithm to effectively search for optimal solutions for heat transfer and fluid optimization \cite{Lohan2016,Zimmermann2018}. Using the genetic algorithm, high performing designs were iterated upon in subsequent steps, producing multiple optimal designs for the researcher to choose. As an example of a GAN used in generative design, Oh uses data mining to collect wheel examples to train a GAN and generate unique designs \cite{Oh2019}. The network generates a random set of input variables to influence a topology optimization stage of the network. Through training, the generator network attempts to produce new designs similar to the examples collected through data mining. The discriminator network is then trained using the sampled outputs from the generator network and the data mined examples to determine which are generated and are from data mining. Through many training iterations, the generator network produces designs indistinguishable from the database. Examples of the generated designs from this network are shown in Figure \ref{topopt3}.

The examples provided only present current research incorporating machine learning in TO and is not an exhaustive review of all applications of TO. A general review of topology optimization advancements for additive manufacturing can be found in Liu et al\cite{Liu2018}.


\subsection{Machine Learning Assisted Modeling of Additive Manufacturing}
As previously discussed, the design space of AM experiments is often vast (e.g., Fig. \ref{AMgene}). While the design of process parameters is often integral to the material design methods reviewed in the previous sections, there are some additional process-centric engineering objectives where machine learning methods may also be beneficial. This section reviews the use of machine learning algorithms to aid in computational design of additive manufacturing process developments. Martukanitz et al. published a full ICME investigation of AM \cite{Martukanitz2014}. There are two modeling scenarios that plague the advancement of AM: the case where a model exists but current numerical methods are too expensive to simulate the model; or the case where a model does not exist. Put slightly differently, in either case $\mathbf{y} = f(\mathbf{x})$ exists but cannot be computed in a reasonable amount of time; or $\mathbf{y} = f(\mathbf{x})$ does not exist. 

Machine learning algorithms have addressed both these cases. In the first case, ML algorithms provide an alternative numerical method for calculating $\mathbf{y} = f(\mathbf{x})$ based on experimental measurements of $\mathbf{y}$ and $\mathbf{x}$, or based on the results of previously run simulations. Machine learning algorithms have also been developed to help visualize trends in high dimensional spaces, allowing researchers to study complex relationships and ask deeper questions. For the second case, ML algorithms provide a form of $\mathbf{y} = f(\mathbf{x})$ from observations (measurements) of the relationship between $\mathbf{y}$ and $\mathbf{x}$.

\subsubsection{Machine Learning as Numerical Methods for Modeling}
There is a suite of numerical methods that have been adopted by the materials science community for computing models of material phenomena. Finite element methods are some of the most common methods. Feedstock, heat source, and melt pool dynamics have been modeled by finite element methods \cite{Toyserkani2004, Khairallah2016, Manvatkar2014} or finite  volume methods \cite{Dai2014}. Some AM-specific tweaks to the finite element method have been developed, such as the quiet/inactive method of Michaeleris et al \cite{Michaleris2014}. They have also been applied to microstructure development \cite{Nie2014}. A review of finite element methods for AM can be found at \cite{Gouge2018}. Models of grain growth in AM have been solved using both phase field numerical methods \cite{Chen2002, Gong2015, Kundin2015, Sahoo2016}, and cellular automata \cite{Tan2011}. Francois et al. provide a review of ICME approaches across spatiotemporal scales\cite{Francois2017}. 

The success of these numerical methods have been in solving complex thermomechanical problems for engineering application. In AM, the number of models that need to be simultaneously considered/computed and the scale of the manufacturing process causes the computational complexity of these methods to grow quickly. Machine learning can make the modeling process more efficient through three primary applications:
\begin{itemize}
	\item Determine a priori which models \textbf{not} to run.
	\item Reduce dimensionality by discarding inputs or physics that are not relevant or that do not have an appreciable impact. 
	\item Compute the same relationship using ML as the numerical method, instead of using explicit methods like FEA, cellular automata, etc.
\end{itemize}

In the case that models \textit{can} be run, but are time-intensive, it behooves the researcher to run as few models as necessary to understand the material response. ML can identify which models will produce the most useful information, informing model choice to the researcher. In another case, ML can be used to identify physics that do not significantly impact the outcome of a model. Reduced-physics models can then be created with a reduced computational burden. In some cases ML can compute the same result as the explicit model with significantly less computational cost (under certain conditions and assumptions).

Dimensionality reduction algorithms identify which parameters are relevant to model in an ICME approach and which are not, enabling future ICME investigations to achieve the same result faster. Materials science has long had a need for dimensionally reduced, computationally accurate models. Some of the first applications of machine learning in materials science was for dimensionality reduction . Dimensionality reduction has been applied to find process-structure-property relations across multiple material length scales \cite{Fischer2006, Flores-Livas2017, Rupp2011, Snyder2012}. Homer applied dimensionality reduction to relate the impact of local atomic environments on mesoscale properties like atomic mobility at grain boundaries, demonstrating the benefit of the technique for advancing ICME \cite{Homer2019}. 

Statistically driven approaches can focus on the parameters in $\mathbf{x}$ that strongly impact AM model outputs, leading to a dynamically guided design of experiments. In design of experiments, a random forest is trained on previously completed experiments. These are the rows of the matrix $\mathbf{B}$ in Eqn. \ref{Bmatrix}. Then, new points in the design space that have not been observed are given to the algorithm and predictions about the output are made. Dimensionality reduction using random forests proceeds differently. 

The random forest is trained on subsets of the data in $\mathbf{B}$. The importance of a feature in the dataset is tested by randomly shuffling the values of one of the columns of $\mathbf{B}$. If randomly shuffling the values of a given feature does not significantly impact the prediction accuracy of the random forest then it is likely that the feature is not important. Towards Data Science provides a more in-depth tutorial on using random forests for feature importance determination \cite{FeatureImportance}. This approach can be applied in computational models that consider many different physics. Several models are run under different initial conditions in the design space. The entires of the matrix $\mathbf{B}$ are the inputs and predicted outcomes of the model. If the exclusion of a model input does not significantly impact the random forest's prediction of the model output, then that input can likely be ignored in future simulations, saving computational time and reducing the number of models that need to be run.

Kamath used a random forest algorithm to screen out irrelevant modeling parameters for predicting maximum density of additively manufactured parts \cite{Kamath2016}. Kamath started with an experimental dataset of manufacturing parameters and multiple modeling methods. An Eagar-Tsai simulation of a Gaussian laser beam on a powder bed was used to model thermal conduction during manufacture, as well as the computationally more expensive Verhaeghe model. The Eagar-Tsai model originally began with four inputs (laser power, speed, beam size, and powder absorptivity) and a design space of 462 possible input combinations. Kamath used random forests to determine which input was most important for achieving fully dense parts. If simulations are time-intensive to run then 462 different simulations may be out of the question. The computational dataset was complemented with an experimental dataset of measured melt pool widths at various printing conditions. Identifying which parameters do not impact the final result reduces the size of input combinations, therefore reducing the number of computations or experiments to be performed. 

Kamath identified that laser speed and power were the most important inputs out of the four to determine melt pool depth and shape. Now that the important physics have been identified, the researchers can proceed to the more expensive Verhaeghe model with knowledge of what parameters to vary.  After determining the most important inputs, the same regression tree was applied to find optimized manufacturing conditions for fully dense parts. However, instead of identifying which features impacted the model standard deviation, the machine settings that maximized $y$ were found. 

A final technique for reducing the burden of computational models requires expressing model data in a matrix and performing matrix factorization. As before, model inputs can be formed into a matrix, $\mathbf{X}$, whose rows are coordinates in the design space. Matrix factorization techniques represent correlations in large datasets in a simplified way. The matrix $\mathbf{X}^T\mathbf{X}$ is a measure of covariance within $\mathbf{X}$. The matrix $\mathbf{X}^T\mathbf{X}$ can be very large due to the design space of additive manufacturing. One type of matrix factorization, called Principal Component Analysis (PCA) represents the data matrix $\mathbf{X}$ as
\eqn
	\mathbf{X} = \mathbf{U} \mathbf{\Sigma}\mathbf{V}^T
	\label{PCA}
\equ
where the columns of $\mathbf{U}$ are the eigenvectors of $\mathbf{X}\mathbf{X}^T$ and are called the \textit{principal components of $\mathbf{X}$}. Similarly, the columns of $\mathbf{V}$ are the eigenvectors of $\mathbf{X}^T\mathbf{X}$. The matrix $\mathbf{\Sigma}$ is a diagonal matrix whose entries are the singular values of $\mathbf{X}$. The first singular value, which corresponds to the first column vector in $\mathbf{V}$, has the highest variance (most information); the second singular value, the second most; and so on. Therefore, regression can be performed on one, a few, or many of the principal components to predict new model results using considerably less data than present in $\mathbf{X}$, but with a minimal loss of information and a minimal reduction in model accuracy.  Materials science studies have used PCA previously to re-represent large datasets in simpler forms, such as predicting the formation energies of crystal structures from a lower dimensional space \cite{Curtarolo2003}. A review of applications of PCA in materials science can be found at \cite{Rajan2009}.

In additive manufacturing, PCA can serve to reduce the number of features in the design space. The new vectors can then be used as regression model inputs for prediction of material properties based on trends observed in dataset $\mathbf{X}$.

\subsubsection{Machine Learning for Visualizing Trends in the Design Space}\label{viz}

Visualizing relationships across high dimensional spaces helps researchers develop an intuitive understanding of data relationships that exist, an intuition that helps guide data preprocessing, feature engineering, model selection, and model training. However, visualizing an $n$-dimensional distribution is difficult. Process maps are commonly employed in AM to visualize 2D slices of the $n$-dimensional AM design space \cite{Beuth2001}. The Ashby plot is a well known generalization of process maps in materials science. Ashby plots show material properties as functions of two design coordinates, such as plotting mechanical strength of various alloys as a function of density and cost to produce. The process maps in welding and AM are more specific versions of Ashby plots. Process maps chart the possible values of machine inputs and identify regions of the design space with similar properties. A commonly employed process map in AM of Ti-6Al-4V describes grain morphology as a function of solidification velocity $R$ and temperature gradient $G$ \cite{DeHoff2015}. Extending process maps to $n$ many process variables would require ${n \choose 2}$ plots. Defining and examining metrics of similarity in an $n$ dimensional space can reveal trends in a human interpretable way without relying on multiple 2D process maps. 

\begin{figure*}
	\centering
	\begin{subfigure}{0.49\textwidth}
		\includegraphics[width=1\linewidth]{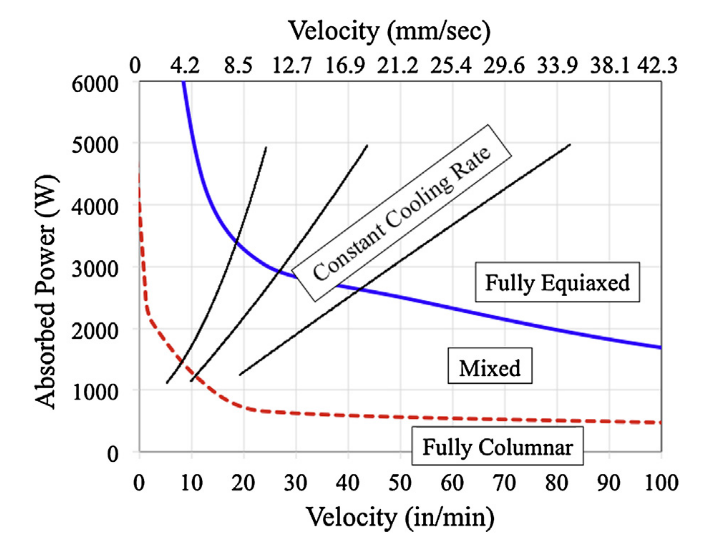}
		\caption{}
		\label{ProcessMap}
	\end{subfigure}
	\begin{subfigure}{0.49\textwidth}
		\includegraphics[width=0.75\linewidth]{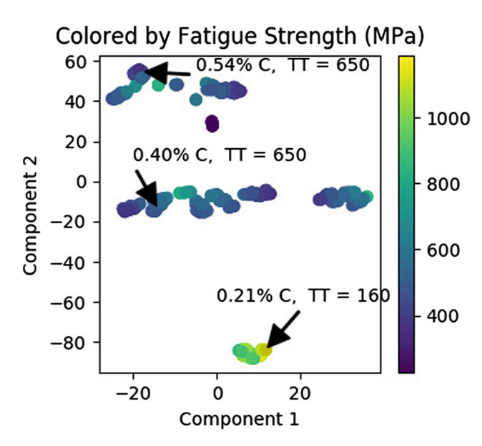}
		\caption{}
		\label{tSNE}
	\end{subfigure}
	\caption{Comparison of a traditional process map and tSNE plot. \ref{ProcessMap}) A process map for predicting microstructure characteristics based on absorbed power and deposition velocity in electron beam wire feed additive manufacturing. Image reproduced from Gocket et al\cite{Gockel2014}. \ref{tSNE}) A tSNE plot from Ling's study showing clusters of samples with similar fatigue strengths\cite{Ling2017a}. While process maps can be useful for predicting manufacturing outcomes they are limited by only showing the behavior of two process parameters at a time. The tSNE algorithm can cluster data based on many manufacturing inputs simultaneously and then display that information in a 2D plot, allowing engineers to study how processing parameters lead to good or bad material properties.}
	\label{ProcessMaptSNE}
\end{figure*}

$t$-distributed Stochastic Neighborhood Embedding (tSNE) is a visualization technique that measures distances in a high dimensional space and then projects data points onto a two dimensional plot. The similarity of all data points in the design space with each other is used to fit a distribution of similarities. The tSNE algorithm begins by fitting a probability distribution to all $\mathbf{x}$'s contained in a dataset. Relationships in $n$ dimensional space are assessed through a \textit{kernel function} $\kappa(\mathbf{x},\mathbf{x'})$ that measures similarity between points in the design space. A commonly employed kernel is the Gaussian kernel
\eqn
	\kappa(\mathbf{x},\mathbf{x'}) = \frac{1}{\sqrt{2\pi\sigma^2}}\exp{\left[- \frac{\mathbf{x} - \mathbf{x'}}{2\sigma^2}\right]}
	\label{gausskernel}
\equ 
where $\sigma$ is a user-specified or fit standard deviation in the distribution of points in the design space. This kernel function assesses distance in the $n$ dimensional space and assigns a similarity value between $\left[0, 1/\sqrt{2\pi\sigma^2}\right]$. 

 After the $n$ dimensional dataset is fit, then a 2 dimensional coordinate $\mathbf{x}^*$ is assigned to each $\mathbf{x}$. The reason for choosing a 2 dimensional coordinate is so that the final result can be visualized on a 2D plot. The tSNE algorithm fits a probability distribution to the $n$ dimensional data set first, then assigns values to each $\mathbf{x}^*$ such that they have the same probability as the associated high-dimensional $\mathbf{x}$. Once the probability distributions have been assigned, the $\mathbf{x}^*$ values can be visualized on a 2D plot to investigate trends.

The benefit of tSNE is that points that are close together in the $n$ dimensional space appear close together on the 2 dimensional plot. This gives AM modelers an idea of how machine inputs and material behavior are distributed in the $n$ dimensional space through a 2 dimensional visualization. Traditional process maps provide similar input/output relationships but are limited in the amount of processing parameters they can interpret at once. A comparison of process maps and tSNE is shown in Figure \ref{ProcessMaptSNE}.

\subsubsection{Machine Learning to Create Models of Additive Manufacturing Processes}
Another problem, equally important to solving models, is the creation of models for additive manufacturing problems. Scientists cannot engineer the additive manufacturing process without an understanding of how process parameters (inputs) impact material properties and performance (outputs). The generation of models in AM is a difficult task due to the large amount of physics that can be incorporated.

Many traditional material models from science and engineering have been applied to additive manufacturing, including thermal history models of heat transfer through the part \cite{Michaleris2014}, residual stress build up during manufacturing \cite{Pal2014, Ding2011}, and thermal signatures such as cooling rate and temperature gradient \cite{Li2014, Raghavan2016}. King et al. provide a review of the physics of AM modeling \cite{King2015a}. 

Phenomenon that are difficult to study experimentally, such as flow within the melt pool, are best studied through modeling approaches. Though expensive, full-physics modeling is often necessary to understand how physics at different scales interact to impact the AM process. If the computational expense of a simulation is too high then performing simulations at all relevant manufacturing conditions can be infeasible. While it is useful for optimization and visualization, reduced order models are unlikely to capture the full dynamics of solidification in AM. 

Within the context of the literature reviewed herein, a \textit{surrogate model} is a regression model that estimates the results of high-cost simulations. Surrogate models are regressed on the inputs and results of previously run simulations. Then, the surrogate model interpolates simulation results at new coordinates in the design space. Surrogate models preclude the need for running computationally expensive simulations for every possible manufacturing condition. Formulating surrogates can be as simple as performing linear regression between simulation inputs and results, but are often more complex. The accuracy of a surrogate model is dependent upon how many previous simulations have been run and at how many different points in the design space.

Tapia et al. built a surrogate model for laser powder bed fusion of 316L stainless steel. They were concerned with predicting the melt pool depth of single-track prints solely from the laser power, velocity, and spot size \cite{Tapia2017}. The dataset used to build the surrogate was computationally derived, based on previous simulation methods used by the same research team \cite{King2014}. In particular, they used the results from a computationally expensive but high-accuracy melt pool flow model of Khairallah et al. \cite{Khairallah2016}. They ran powder bed simulations at various laser powers, velocities, and spot sizes, and the model told them the depth of the melt pool, amongst other information. The datasets provided enough information for a surrogate model to be trained to predict simulation results.

\begin{figure*}
	\centering
	\begin{subfigure}{1\textwidth}
		\includegraphics[width=1\linewidth]{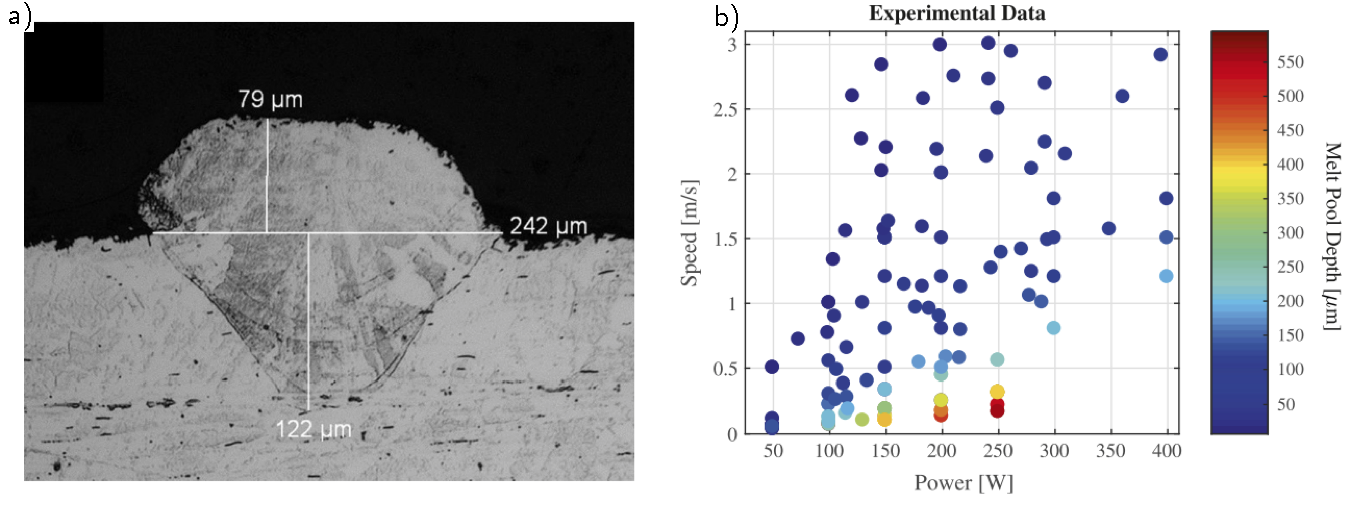}
		\label{InputData}
	\end{subfigure}
	\begin{subfigure}{1\textwidth}
		\includegraphics[width=1\linewidth]{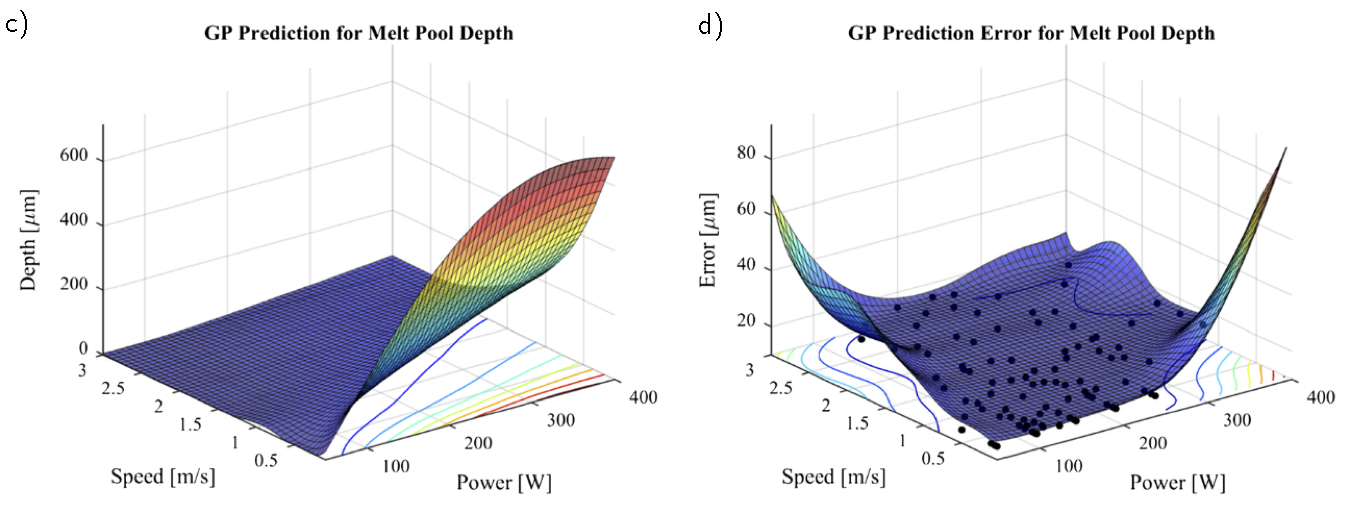}
		\label{ResponseSurface}
	\end{subfigure}
	\caption{Input data, response surface $y(\mathbf{x})$, and error $\epsilon(\mathbf{x})$ of Tapia's study in predicting melt pool depth for laser powder bed fusion \cite{Tapia2017}. a-b) The input data used to train a Gaussian Process Regression model for predicting melt pool shape and depth. They started with a sparse sampling of the design space to build the model. c-d) The Gaussian process model predictions for melt pool depth as a function of two input parameters $y(\mathbf{x})$ (left) and the associated prediction errors $\epsilon(\mathbf{x})$.}
	\label{GPMs}
\end{figure*}

To build their surrogate model, Tapia used a machine learning algorithm known as a Gaussian process model (GPM). A common model assumption in Gaussian process modeling is
\begin{equation}
	z(\mathbf{x}) = y(\mathbf{x}) + \epsilon(\mathbf{x})
	\label{model}
\end{equation}
where $y(\mathbf{x})$ is the approximation (surrogate) of the simulation process, $\epsilon(\mathbf{x})$ is a stochastic, randomly distributed noise in measurement, and $z(\mathbf{x})$ is the value given by a simulation. The primary goal in GPMs is to find model parameters for the mean process $y(\mathbf{x})$ and a covariance function $\kappa(\mathbf{x},\mathbf{x}')$, which is a function of similar form to Eqn. \ref{gausskernel}. Fitting a Gaussian process model often begins with assuming a model function for covariance, fitting the model parameters such as $\sigma$ to the observed values $z(\mathbf{x})$, then using those model parameters to predict simulation results $y(\mathbf{x})$ at other locations in the design space. 

Tapia used Bayesian statistics to develop a probabilistic model that predicted melt pool depth from simulation inputs. They were able to successfully predict the outcomes of both high-fidelity simulations and experimental measurements solely by analyzing trends in previously obtained results. In particular, they were able to accurately predict the melt pool depth at a value that had never been observed before, either computationally or experimentally. For future investigations, predictions by the surrogate model can be relied upon instead of running a simulation or experiment. Regression models such as this provide engineers with faster routes toward optimized manufacturing states by predicting manufacturing at a wide range in the design space based on only a few initial experiments.

Gaussian Process Models provide robust uncertainty metrics on the predictions they make. Uncertainty estimation is important in materials informatics because it enables scientists to know how confident their models are in predictions in various regions of parameter space. Some machine learning models do not have straightforward ways of assessing model error \cite{Bessa2017}. 

Another benefit of GPM is that it aids in inverse design and design space visualization. GPMs can explicitly identify regions of the design space that will maximize or minimize a value. In the case of Tapia et al. response surfaces were created from the GPM that visualized the depth of melt pools as a function of laser power and speed. Doing so allows engineers to identify regions of the design space that provide specific material responses, an important tool in optimization for additive.

Machine learning is not only limited to ex situ experimental investigations or modeling approaches. Ideally, machine learning can be used to solve multiobjective optimization functions where multiple aspects of the AM process are optimized at once -- energy density, melt pool shape, heat transfer, grain growth, and the list continues. Models can be created that solve this multiobjective optimization problem and present to the engineer what an optimal manufacturing process looks like. Actually \textit{creating} that optimal process will require tight control of the manufacturing process. Machine learning models trained on correlations between build parameters, the dynamic response of the system, and the final part properties can be combined with real-time computer vision to simultaneously observe, characterize, and control many different aspects of the manufacturing process.

\subsection{Process Monitoring and Control}

The numerousness of signals to measure in situ for AM warrants use of quick, efficient, and robust signal processing methods for process monitoring, feedback, and control. These signal processing algorithms are closely related to machine learning. They serve as tools in their own right, and can also pre-process data for use in other machine learning applications, like clustering and regression. Computer vision is one class of image recognition algorithms that has been developed for automated feature identification in signals. Intelligent computer vision uses ML algorithms to identify objects and features in a wide variety of data types. We proceed to discuss the potential uses of data analytics and ML to advance our ability to directly study and control AM process technologies.

\subsubsection{In Situ Process Monitoring and Feedback}\label{sec:In Situ Process Monitoring and Feedback}
\begin{figure*}
	\includegraphics[width=0.85\linewidth]{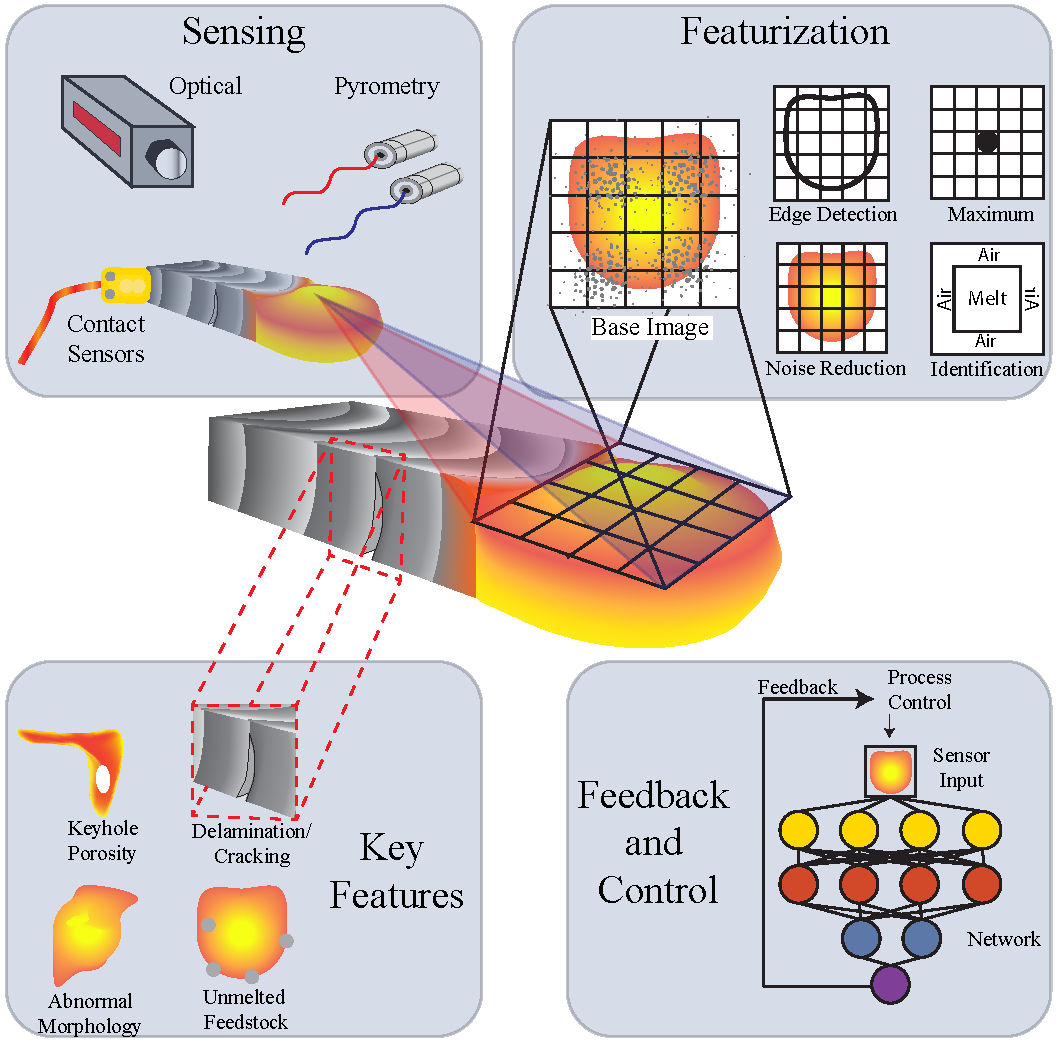}
	\caption{A few examples of data types, data sensors, and features to detect in a laser powder bed fusion manufacturing process. The wide range of signals to monitor then control makes feedback and control in AM especially difficult. Computer vision techniques can be applied to automatically detect features of interest across multiple data types and data sensor simultaneously.}
	\label{fig:melt_pool}
\end{figure*}

In situ monitoring, feedback, and control has been consistently ranked as one of the most-needed technologies for advancing additive manufacturing \cite{Berumen2010, Tapia2014, Mani2017}. The combination of rapid solidification and the small length scales of AM solidification can make traditional process monitoring approaches difficult. Furthermore, there are many processes/problems to monitor for during the manufacturing process, with equally as many sensor types for monitoring as shown in Figure \ref{fig:melt_pool}. Machine learning can fill in gaps that leverage correlations and relationships from previous measurements, observations, and responses.

Process monitoring involves acquisition of real-time signals that can reveal information about a wide variety of phenomenon during manufacturing. Many developments of in situ process monitoring technologies are focused on controlling a) microstructure growth or development; or b) the prevention of defect formation. 

There are in-situ experiments being performed to inform models of the additive manufacturing process. In situ experiments advance our understanding of AM, as well as advance feedback and control for AM, through several outcomes. In some cases,  in situ studies reveal what a `good' or `bad' AM process looks like. They also inform researchers of those conditions that must be met to achieve a desired outcome or prevent the formation of a defect. In situ experiments also push the development of sensor technology for AM. While sensor technology will not be covered in this review it is an important topic for the advancement of AM technology. Purtonen et al. wrote a review of common sensing methods for laser based manufacturing\cite{Purtonen2014}.

Early experiments using in situ monitoring for AM focused around either the ability to measure thermal signatures accurately or relating key features of the solidification process to important material properties. McKeown et al. has used dynamic transmission electron microscopy to measure solidification rates in powder bed AM \cite{McKeown2016}. Bertoli et al. have also characterized cooling rates using high speed imaging \cite{Bertoli2017}. Raplee et al. have used thermography to monitor the solidification and cooling rates of electron beam powder bed fusion, relating the temperature profiles to defect and microstructural characteristics \cite{Raplee2017}. Distortion of parts due to thermal cycling was investigated by Denlinger et al. by means of thermocouples in contact with the build substrate \cite{Denlinger2015}. Guo et al. used synchrotron X-ray imaging to characterize the dynamic behavior of spatter during laser-based AM \cite{Guo2018}. Leung et al. likewise used synchrotron X-ray imaging to characterize defect formation and molten pool dynamics during laser powder bed fusion \cite{Leung2018}. Based on the behavior they observed, Guo et al. were able to suggest control mechanisms for minimizing spatter during manufacture. Everton et al. provide a review of in situ monitoring for metal AM \cite{Everton2016}. All of the data being recorded in these studies can be used as \textit{features} for training machine-learning based feedback and control systems. The class of algorithms used in these cases is called computer vision.

The type of data being collected in situ is often in the form of time series or image data. In computer vision, as with traditional feedback and control, algorithms are used to identify deviations from a desired signal. The power of computer vision approaches is their ability to simultaneously monitor and identify signal changes across multiple sensor types, as well as multiple different types of deviation from a single sensor. Examples include identifying a spike in temperature or a sharp change in intensity in an image indicating a deviation from a desired processing conditions. Image processing \textit{filters} can be used to selectively modify or extract features in AM data. Image processing filters are mathematically analogous to those introduced for topology optimization (Section~\ref{sec:topology optimization}). 

A filter is implemented as a mathematical operation, a kernel, applied to a window of time series data or an area of pixels in an image. For images, as previously discussed in Section \ref{feat}, filters attempt to use local spatial information and \textit{a priori} knowledge of the expected properties of the image to improve image quality and extract features, e.g., distinctive characteristics such as edges or regions of similar intensity (domains) that represent the boundaries or spatial extents of objects, phases, etc.

AM processes span several orders-of-magnitude in both length and time scales from ejected particles moving across the field of view in milliseconds to multi-hour builds and sub-millimeter melt pools to part-scale thermal distortions. Practically, then, in situ monitoring requires compromises in data collection rates and resolutions, and data processing filters are used to reduce noise and extract features, such as melt pool width, from the as-collected data. A comprehensive review of image filters is beyond the scope of this review, so the interested reader is directed to the many works on this topic, such as Vernon et al\cite{Vernon1991}. However, three use cases are especially common and worth discussion here: reduction of high-frequency noise, also known as salt-and-pepper noise; additive noise reduction; and edge detection.

High frequency noise is characterized by sudden changes in intensity relative to the surrounding field. Although there are a number of possible causes, this may be caused by pixel-level variability or insufficiency in the detector, e.g. ``dead pixels'' or excessive gain. Median and conservative filters are commonly used when the fraction of noise pixels is large (1\%--10\%) and small ($<$ 1\%), respectively.

Additive noise, unlike high frequency noise, is a result of insufficient counting statistics, which may result from insufficient exposure time, or detector efficiency. A gaussian filter adjusts the intensity of each pixel according to the weighted intensities of neighboring pixels. Unlike median and conservative filters, a gaussian filter will soften edges, making adjacent domains less distinct.

Filters also have applications beyond noise reduction, primarily in object and feature detection. Detecting phenomena of interest during manufacturing is the first step to feedback and control mechanisms. Edge detection captures local changes in intensity to identify transitions between adjacent domains. Laplacian or Laplacian of Gaussian (LoG) filters themselves are sensitive to noisy images, identifying spurious edge artifacts, but are used as part of larger algorithms, such as Canny edge detection~\cite{Canny1986}. Canny edge detection include noise reduction to mitigate artifacts of LoG filters, and can be used to monitor melt pool shape and identify other features, such as unmelted powder particles attached to the build surface. Canny edge detection, along with other feature extraction algorithms, can be used to extract the features that characterize the build and can be used as part of a larger machine learning workflow to classify build quality. For example, these features can be used in learning algorithms to correlate characteristic features, such as melt pool width and hatch spacing, with particular behaviors, such as the formation of lack of fusion defects, in the manufacturing process. In this case, identification of a feature, or set of features, may be sufficient to indicate a particular process outcome.

Template matching is a computer vision method that can be used for automatic identification of common patterns. It involves the comparison of an unclassified input to a database of pre-identified patterns. For AM, template features include abnormal melt pool morphologies \cite{Kanko2016}, inclusion of unmelted powder particles \cite{Yang2017}, and denudation near the melt zone \cite{Matthews2016}. The scale-invariant feature transform (SIFT) \cite{Lowe2004} and a variant, ``Speeded-Up Robust Features'' (SURF) \cite{Bay2008} are both feature identification algorithms that can be used for template matching. Another template matching algorithm is the \textit{bag of visual words} or dictionary method~\cite{DeCost2015}. A collection (dictionary) of typical features from the AM process can be built based on features obtained from filters. The features measured in situ are compared with dictionary entries. If an in situ feature matches a defect-indicative feature from the dictionary, then it is likely a defect has formed during manufacturing.

\begin{figure}
	\includegraphics[width=0.48\textwidth]{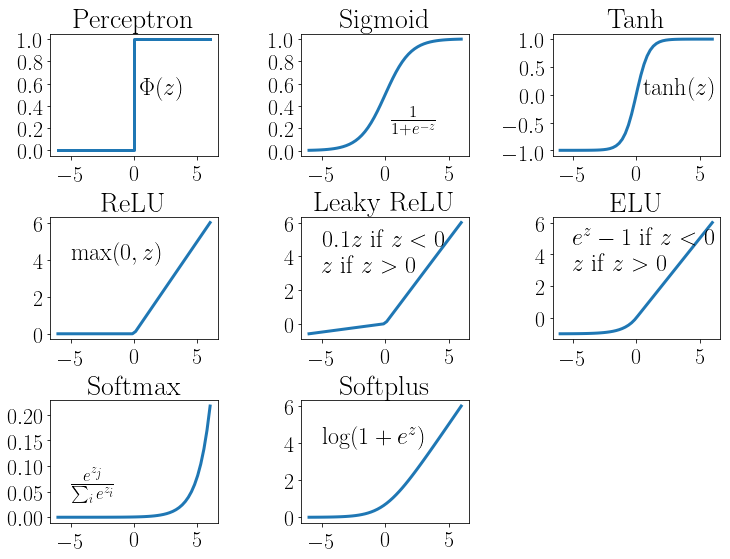}
	\caption{Common activation functions in artificial neural networks (NNs) that introduce nonlinearity into the NN. The sigmoid is the archetype activation function because the closed form solution for the derivative of the sigmoid, which is used during model fitting, is an excellent pedagogical tool; however, the rectified linear unit (ReLU) is, at present, the most common activation function in the hidden layers of NN. Uses for the other activation functions are provided in the text.}
	\label{fig:activation functions}
\end{figure}

Neural networks (NNs) are particularly well-suited to handle features extracted from images, or simply the images themselves. There are many references that describe neural networks in detail, such as the work of Hastie et al\cite{Hastie2009}, and an increasing number that address the specific challenges associated with neural networks in materials science~\cite{Bhadeshia2009}. There are several properties of NNs that are worth repeating here, however. Each layer in a NN is connected to the next layer through an affine (linear) transformation. This step stretches, scales, and skews the input vector.
\begin{equation}
	{\bold z}^{(i+1)} = \boldsymbol \theta_i^T {\bold x}^{(i)}
\end{equation}
where ${\bold z}^{(i+1)}$ is the input into the $(i+1)$ layer and ${\bold x}^{(i)}$ is the output from the previous, $i^\textrm{th}$ layer. Then, an activation function, such as those summarized in Figure~\ref{fig:activation functions}, introduces a non-linearity that warps/distorts the vector input to that layer.

\begin{equation}
	{\bold x}^{(i+1)} = f \left( {\bold z}^{(i+1)} \right)
\end{equation}

The model parameters $\mathbf{\theta}_i^T$ are regression weights that associate outputs from each layer $\mathbf{x}^{(i)}$ to subsequent layers $\mathbf{z}^{(i + 1)}$. By increasing the depth of the NN, that is, adding additional layers, and the width (number of nodes) of those layers, a NN can be used to approximate any function, making them powerful regression and classification tools~\cite{Hornik1989}. However, the general sparsity of materials data coupled to the complexity of process--structure--process relationship requires an understanding of the tradeoffs and requirements of using NNs in materials science, and in AM more specifically. Beyond the basics of model architecture, overfitting and the bias--variance tradeoff that is part of any machine learning model, a basic understanding of the role of activation functions can help to develop an intuition for the use of NN in materials and manufacturing.

An early use of NNs was in classification. The perceptron, logistic sigmoid (or simply, sigmoid), and hyperbolic tangent are all activation functions that choose between two options (0 or 1, or in the case of $\tanh$, -1 or 1). While a binary option may seem overly limiting, even multinomial classification can be broken down into a sequence of such binary classificiations: \textit{A} or not \textit{A}; and if not \textit{A}, then \textit{B} or not \textit{B}; and if not \textit{B}, \textit{C} or not \textit{C}; etc. However, such a serial solution will require more layers and, with more layers, longer training on larger datasets to fit all model parameters.

Visual examples of these activation functions can be seen in Figure \ref{fig:activation functions}. While each behaves differently, particularly across the negative domain ($x < 0$), the simplicity and robustness of the ReLU have made it the most commonly used activation function for hidden layers in regression neural networks.

In the case of a multinomial classification problem, a more simple network may be possible by using one-hot encoding. A one-hot encoding vector is defined for $N$ exclusive options: one element in the $N$-element vector is 1, all other values are 0. Rather than using multiple layers to construct the binomial ladder required to simulate a multinomial decision, the softmax activation function selects one-from-many in a single layer. Since each value in the input vector appears in the softmax exponent, even small differences in the magnitude of $z$ result in large differences in the output of this activation function; therefore one option, represented by one node or neuron in the layer, is approximately 1 and all others are nearly 0. Simplification of the network architecture by choosing activation functions that more closely resemble the nature of the problem emphasizes the importance of domain-specific knowledge in developing appropriate NN architectures.

Combining the concepts of neural networks and image processing filters, convolutional neural networks (CNNs) not only learn how to correlate features to results, they are designed to also identify the filters that extract those features. These networks require large numbers of parameters, in the tens to hundreds of millions, that introduces an insurmountable training burden due to the sparsity of materials data. However, CNNs trained on natural images have demonstrated a remarkable similarity in their initial layers~\cite{Yosinski2014}. These first few layers identify basic shapes, edges, and colors that are common to many image types; a phenomenon that many groups have exploited to overcome the limitation of data sparsity through transfer learning~\cite{Ling2017a}, including specific work in the field of additive manufacturing. Yuan et al \cite{Yuan2018} were able to successfully monitor melt track width, standard deviation, and continuity of tracks in situ during laser powder bed manufacturing. Scime and Beuth trained a convolutional neural network to identify six different types of defect that are typical of laser powder bed fusion, with reasonable prediction accuracy \cite{Scime2018}. Li et al. used a type of neural network method called \textit{deep learning} to classify AM parts using microstructural images \cite{Li2020}. Kwon et al. classified melt pool morphologies using a neural network \cite{Kwon2018}. These studies represent only a few possible uses of CNN for in situ process monitoring of AM.

Scime and Beuth modified a well-known convolutional neural network architecture -- known as AlexNet \cite{Krizhevsky2012} -- to perform classification of powder spreading errors that occur in laser powder bed fusion \cite{Scime2019}. The study presented by Scime and Beuth go in-depth on the architecture of their CNN and directly explain how the training of filters applies in the context of AM images. 

\subsubsection{Featurization of Qualitative Image Data}
The same processing algorithms that are used for featurization and modeling of in situ signals can also be applied to automate part of the scientific process of studying additive manufacturing. Specifically, computer vision can be used to automate classification of microstructures during parametric analysis.

Parametric analysis in additive manufacturing requires the characterization and measurement of material properties that result from a specific coordinate in the design space. Often, material properties like mechanical strength, surface roughness, microstructure, or defect density have to be measured, analyzed, and quantified or classified as part of the parametric analysis process. This experimental process can be tedious. More often than not, images are relied upon heavily in classifying material properties, especially microstructures. Fortunately, machine learning algorithms can be applied to automate the analysis of images during the parametric analysis process.

It is worthwhile to mention up front that these algorithms have been \textit{tested} on microstructure and, in some cases, additive-specific images. There are few algorithms that can process AM microstructure data `out-of-the-box.' Rather, these algorithms will need to be tailored in order to quantify AM images specifically. However, the algorithms discussed here have been proven on non-AM microstructure datasets, thus they should be extensible to AM datasets. The computer vision approaches that work for microstructure data are often the same approaches that will be discussed again later for in situ monitoring and feedback.

One AM-related application of image characterization is measuring particle size distributions in AM powder feedstock. DeCost and Holm used SIFT with a dictionary classifier to measure the particle size distribution for a dataset of synthetic powder particles \cite{DeCost2017a}. Particle size distribution plays in several steps across the additive process including energy absorption and part metrology \cite{Zhou2009, Boley2015, Boley2016}. DeCost created datasets with six different particle size distributions. Image features were identified and classified using $k$-means clustering on the features found by SIFT. Then, a classification algorithm known as a support vector machine (SVM) was trained to classify image features into particle sizes. DeCost was able to achieve $89$\% overall classification accuracy in measuring particle size distribution this way. DeCost later improved upon this powder classification method and were able to achieve higher classification accuracies for real powder images \cite{DeCost2017}. Machine learning algorithms have also been trained for the automatic classification/identification of EBSD texture maps \cite{Shen2019, Kaufmann2020}.

\begin{figure*}
	\includegraphics[width=1\linewidth]{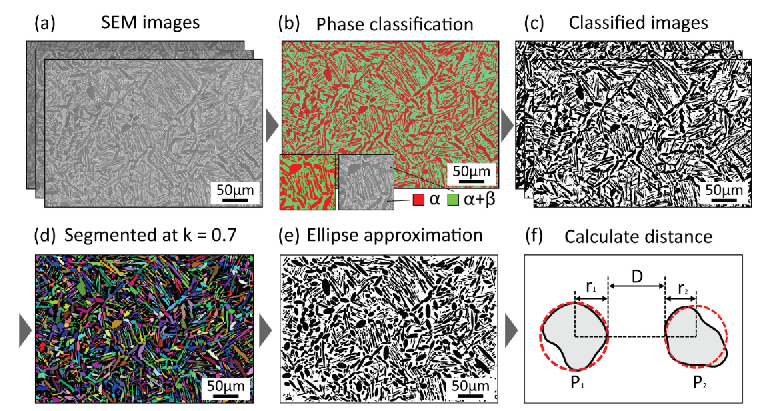}
	\caption{The image segmentation approach implemented by Miyazaki et al. to automatically segment, classify, and characterize SEM images of Ti-6Al-4V microstructures. a) First, the SEM images are obtained. b) A random forest algorithm is used to classify regions of the image and c) build a database of classified images. d) Image segmentation proceeds to separate out the $\alpha$ and $\beta$ phases. e) An ellipse approximation is overlaid on the segmented image to characterize grain morphology and size. f) The nearest neighbor distance can be calculated from the ellipse locations to provide a measure of grain distribution. Microstructures can be very complex for additively manufactured alloys and performing this characterization by hand becomes burdensome. Image recognition algorithms can automate the process and significantly speed up characterization, development, and qualification times. Image reproduced from Miyazaki et al\cite{Miyazaki2019}.}
	\label{MiyazakiSegementation}
\end{figure*}

Strides have been made in automatically identifying and quantifying information from metallographs \cite{DeCost2015, DeCost2017b, Ling2017a, Bulgarevich2018}. A good portion of quality control in materials science as a whole, not just AM, involves classifying materials based on metallographs or micrographs of microstructure. Work is being done across materials science to apply machine learning based computer vision to classifying and quantifying information in these microstructural images. Doing so will speed up the process of materials characterization and qualification, while also providing methods of quantifying information that otherwise would have stayed in a qualitative form. Examples include classification of grain structures, measurements of grain size, pore size calculations, and more. 

An additive-specific image segmentation algorithm was used by Miyazaki et al. \cite{Miyazaki2019}. Five image filters were convolved with microstructure images of selective laser melted Ti-6Al-4V. The features identified by these filters were used in a random forest algorithm to segment the image into regions of $\alpha$ phase grains and $\beta$ phase grains. The algorithm was able to automatically calculate area fraction of primary and secondary $\alpha$ phases that form during cooling. It was also able to calculate the nearest-neighbor distance between grains. Nearest neighbor distance of grains is indicative of grain characteristics like size, morphology, and distribution. 

Chowdhury et al. took a more expansive approach to performing feature identification in microstructures. In particular, they were looking to classify microstructures as either dendritic or non-dendritic. Chowdhury employed 8 different feature identification methods for a dataset of images. Classification was performed using an ensemble of ML techniques including support vector machines (SVM), Na\"ive Bayes, nearest neighbor, and a committee of the three previous classification methods \cite{Chowdhury2016}. Chowdhury's wide approach to image classification achieved classification accuracies above 90\%. 

Efforts are underway across materials science to implement computer vision for the automation of materials classification. Rather, the authors would like to refer the reader to the computer vision libraries listed in Table \ref{data_tools}.

\section{Learning from the Past: Moving Towards Database-Driven Design of Additive Technologies}
The scientific approaches to studying additive manufacturing discussed herein -- parametric analysis, computational modeling, in situ monitoring, and the like -- produce data. The application of machine learning to these scientific approaches likewise produces data. All of this data comprises a subset of the AM design space. The integration of this data into multi parameter, multi physics, multi printer datasets increases both the size of the design space that can be explored as well as the depth/accuracy at which certain regions of the design space can be modeled. Making AM process-structure and process-property data open and accessible to the scientific public accelerates the rate at which data-driven approaches can help to advance AM research and engineering. This potential is evident in examining some more mature examples of the use of data-driven approaches in materials science and engineering, which we proceed to briefly review in this section to motivate the development of data-driven approaches for AM. 

Databases of process-structure and process-property relationships are not a new concept in materials science. Databases like the Linus Pauling Files or International Crystal Structure Database have been widely used for materials design. Domain-specific databases are also being generated from high throughput experimental and computational investigations that have occurred over the past thirty years. Experimental high throughput investigations have also been used in materials science for many decades \cite{Xiang1995}. Common deposition techniques (sputter, plasma, vapor, etc.) have enough degrees of freedom to allow for continuous compositional variation within a single sample, which allows for continuous mapping of composition-structure-property relationships \cite{Long2007, Long2009, Kusne2015a}. These combinatorial synthesis methods present analogous design space challenges to AM: the number of possible input combinations obscures many of the important underlying process-structure phenomenon. It has long been established that synthesizing and characterizing large combinatorial catalogues of samples can lead to the discovery of materials with optimized properties faster than a theory-driven approach by itself \cite{Ceder1998, Pilania2013}. High throughput deposition studies with chemical vapor deposition, metallorganic chemical vapor deposition, physical vapor deposition, and atomic layer deposition, among other techniques are commonplace for the manufacturing of sensors, batteries, photovoltaics, electronics, shape memory alloys, and the like \cite{Hampden-Smith1995, Gilmer1998, Mercey1999, Mitzi2001, Cui2006,Dwivedi2008, Jin2013}. Furthermore, the parameters of interest in these studies can sometimes be quickly catalogued using high throughput characterization techniques like laboratory X-ray diffraction and electron probe microanalysis \cite{Gregoire2014, Ren2017}. These combinatorial studies culminate in large libraries of material properties listed as a function of composition. As far back as the 1990s, data-driven algorithms were being applied to search and discover using these large libraries of composition-property data. Evolutionary and genetic algorithms were trained on composition to predict stable crystal structure and material properties \cite{Deaven1995, Morris1996, Woodley1999, Stucke2003, Wolf2000}. Even neural networks, which did not have the widespread use then that they have now, were being applied for the prediction of crystal structures based on composition \cite{Sumpter1996}.

Modeling challenges in materials science have also been tackled using large databases with machine learning. Packages such as the Vienna Ab initio Simulation Package (VASP) have been employed for high throughput searches of stable material systems with a wide range of properties. The stability and maturity of these packages have enabled the reliable automated calculation of new stoichiometries and new phases \cite{Glass2006} and enabled the automated and semi-automated search for new functional materials\cite{Hafner2006}. As these methods have improved, computational high throughput investigations continue to increasingly match and provide complementary information to experimental measurements \cite{Curtarolo2005}.  High throughput density functional theory (DFT) studies generate quite a bit of data and are therefore well equipped for machine learning and database-driven design. The application of high throughput DFT is widespread for design of materials with all sorts of properties including high temperature superconductors \cite{Kolmogorov2006}, lithium ion batteries \cite{Kang2006, Chen2012, Kirklin2013}, molecule design \cite{Mannodi-Kanakkithodi2016, Butler2018}, cathode materials \cite{Hautier2013}, piezoelectrics \cite{Roy2012}, ferroelectrics \cite{Bennett2012}, corrosion resistant films \cite{Ciobanu2005}, and thermoelectrics \cite{Wang2011, Yan2015}. Each of these studies, like parametric studies in additive, vary a set number of model input parameters and measure a material property as the dependent response.

Yet many of the same modeling obstacles exist in DFT as in AM, such as a lack of transferability between models and the computational expense of large material systems. The design space problem exists here as well -- there are so many possible compositional combinations that knowing \textit{where} to look is difficult. Machine learning was proposed as a solution for obstacles in high throughput DFT as early as 2005 \cite{Morgan2005}.  Large unit cells whose properties cannot be directly calculated using DFT are often approximated using machine learning approaches like neural networks \cite{Behler2015}, genetic algorithms \cite{Hart2005}, and principal component analysis \cite{Snyder2012}. Studies applying machine learning to databases of computational information have gone beyond tackling computational problems. In some cases, the studies have revealed previously unobserved or uncharacterized relationships between crystal structure information and materials properties \cite{Ghiringhelli2015}.

In other efforts to reduce the time to design and deploy new materials, programs like the Materials Project incorporate data taken from a wide range of experimental and computational methods into an open-source, accessible database. The Materials Project also features electronic, structural, and thermodynamic calculations of different materials as well as an automated workflow for doing DFT computations of material systems \cite{Jain2011, Jain2013}. Other databases of materials information include AFLOWLib \cite{Curtarolo2012, Curtarolo2012a}, the Harvard Clean Energy Project \cite{Hachmann2011}, Japan's National Institute of Material Science \cite{NIMS}, and the Open Quantum Materials Database \cite{Saal2013}. Some pipelines for high-throughput computation and analysis have included consideration of publication timelines in their processes \cite{Foster2015}. These databases offer a multitude of benefits to materials researchers. First and foremost, publicly accessible databases offer an infrastructure for the free flow of experimental and computational results. Synergy between research groups becomes easier as data is shared more freely. Furthermore, many of these online databases also provide tools for performing material design. The Materials Project offers a design interface, whereby users can specify a set of material properties and are provided with a list of likely candidate materials. Other projects, like AFLOW, allow for fast high-throughput DFT calculations of a wide range of material systems.

The generation of databases that are accessible to the scientific public is a primary step on the roadmap of the Materials Genome Initiative \cite{DePablo2014}. Much of the development of materials databases have focused on computationally-derived materials information. Infrastructure and standards need to be developed that allow for sharing of experimental data that is understandable and usable by many researchers. Data journals are becoming more common for sharing datasets from scientific investigations and are making strides in standardizing data-sharing infrastructure \cite{Wilkinson2016}, along with the publication of datasets themselves for public use \cite{DeJong2015, Kim2017}. By examining the development of image processing databases outside of materials science, it is evident that the collection and distribution of image databases have enabled rapid developments in the field of computer vision. Many of the more common objectives with computer vision -- autonomous navigation, face recognition, object recognition, image segementation -- have databases that are catalogued in online repositories like CVonline \cite{CVonline} and VisionScience \cite{VisionScience}. Learning from these other fields, open sharing of AM microstructure image databases will aid in the development of segmentation and identification algorithms that are suited for materials, and more specifically AM-specific problems.

Having open, accessible databases improves the rate at which machine learning can be applied to design for additive manfuacturing. Machine learning as a tool driving materials design was proposed some time ago. Review articles have explored the many and varied uses of machine learning across materials science, with many of the applications finding great success \cite{Kalidindi2016, Ramprasad2017, Gubernatis2018}. A review article on best practices for machine learning in materials science can be found in the work of Wagner et al\cite{Wagner2016}. Open sharing of databases also tackles a problem in ICME approaches to AM; that is, the \textit{integration} of multiple data sources. AM incorporates relevant physics over many different time and length scales, to the extent that a single research group is unlikely to have access to all pertinent information. Open sharing of data sets, whether it is computationally derived, experimental, or images, allows research groups to incorporate multiple physics simultaneously. Furthermore, it will accelerate the rate at which AM materials research is performed as higher fidelity machine learning models can be built with more and diverse datasets.

Additive manufacturing should move toward the same types of infrastructure for open data sharing. The combinatorial problems in additive are widespread and cover many, many length scales. Large institutions may have the resources to link time- and length-scales in additive manufacturing. Smaller research groups are often limited to studying a single process phenomenon and do not necessarily have means to integrate their knowledge into other additive manufacturing studies. The generation of additive databases allows for a democratization of research and an acceleration of the pace at which additive manufacturing advances are made.

\section{Conclusions}
Materials informatics has demonstrated great success as a tool that can accelerate and reduce cost for discovery, design and optimization of many material systems. Metals additive manufacturing is primed to benefit from the same algorithms and statistical models. Many of the major obstacles that lie ahead in additive manufacturing -- fully integrated ICME modeling approaches, data-driven design, feedback and control using in situ process monitoring sensors -- can be attained by incorporating machine learning. However, machine learning itself is not the end-all-be-all solution to developing AM technologies. There are many obstacles in the application of machine learning itself that will need to be addressed along the way. ML is a complementary tool to physics-based modeling and experiments. Just like transmission electron microscopy doesn't solve every problem by itself, neither will machine learning. Instead, it should be understood where and when ML is a desirable techinque, and which class/type of ML algorithms is right for the problem. Since the goal of this review article is to introduce AM scientists and engineers to the concepts of ML and the selection and evaluation of ML algorithms for solving problems in AM, in conclusion, it is worthwhile to summarize the major AM challenges that can be solved using machine learning, as well as identify the major obstacles to implementation.

\subsection{Key Application Areas for Machine Learning in Additive Manufacturing}
\begin{itemize}
	\item \textbf{Coupled Physics-Statistics Models:} The original goal of materials informatics, dating back to high throughput thin-film studies in the 1990s, was to model material process-structure-property relationships that were highly complicated and lacked a single governing physical theory \cite{Xiang1995}. Additive manufacturing is the embodiment of a complicated physical system, where governing equations across optics, fluid mechanics, solid mechanics, thermodynamics, and kinetics have to be incorporated into one model. Machine learning can build computationally accessible surrogate models of more complicated physical systems that are useful for engineering and design. 
	
	\item \textbf{Materials Design:} Materials design through machine learning has already been applied in a wide range of fields cited here, including thermoelectrics, photovoltaics, semiconductors, Heusler compounds, and many, many more. Design in these fields typically focuses around combinatorial studies of compositions, crystal structures, and a material response. Materials are manufactured through a wide variety of techniques but optimization is rarely applied to the manufacturing method itself, just the materials used in manufacturing. In additive, not only does the material system need to be tailored but the conditions of manufacturing also need optimization. Materials properties to consider range from composition and atomic properties to phase kinetics. Manufacturing optimization includes the energy density used, deposition rate, feedstock supply mechanism, and more. Machine learning can integrate optimization across these separate design considerations. Process optimization is likely to include in situ control.
	
	\item \textbf{Automated Process Control:} There are many variables to monitor and keep track of in the additive processes. There are equally many sensors and measurement techniques to monitor the process. Advancements in signal processing and computer vision must be taken advantage of to build incorporating process control models. Intelligent feedback and control for additive can simultaneously integrate and understand multiple signal types \textbf{and} optimize on multiple objective functions simultaneously. Taking full advantage of the promises of AM -- topologically optimized geometries, functionally graded materials, minimized design-to-fly time -- will require tight control over the manufacturing process.
\end{itemize}

\subsection{Further Developments are Needed in Both Additive Manufacturing and Machine Learning}
\begin{itemize}
	\item \textbf{Data Sharing Infrastructure:} Programs like the Materials Project, AFLOW, and OQMD have accelerated the rate at which materials design can occur, as well as the rate at which scientific data is shared. The democratization of data has allowed many different research teams to search through the materials design space in search of new materials, to great success. The same type of democratization is possible in additive if infrastructure exists for sharing of AM data. However, standardization of AM data types should be addressed before data can be shared in a useful, meaningful way.

	\item \textbf{Curation of Data and AM Standards:} Success in applying data-driven approaches is tied tightly to the quality of data being used. Even data that has been collected with the highest care and precision can be detrimental to a model if it is labeled incorrectly or inconsistently. Work is proceeding in standards development for additive manufacturing \cite{Seifi2016}. However, additive manufacturing technology development has sometimes proceeded faster than standardization. Care needs to be taken in developing AM standards that are consistent across manufacturing devices and can also account for developments in the broader technology.
	
	\item \textbf{Experimental Measurement and Sensor Development:} While in situ measurement devices are widespread, the time and length scales of additive manufacturing can push the limits of current high-end sensors. Imaging methods that can resolve the fast, dynamic, microscale melt pools of additive would allow for a huge leap in process monitoring and control. Equally important is developing methods of determining temperature history throughout the duration of builds. Both of these technologies are crucial for fine control over the additive process. 

	\item \textbf{Physics-Informed-Data-Driven Models:} Additive manufacturing has developed amazingly over the past few decades thanks to traditional scientific and engineering approaches in many different fields. Modeling AM using classical thermal, mechanical and kinetic models has shown success in advancing and engineering the technology. This review is suggesting that machine learning be used as a complementary tool to these traditional approaches. It would be unwise to completely ignore physical theories that have shown applicability in AM. Rather, machine learning algorithms should be built around currently existing models. There are equally rich mathematical frameworks in both materials science and machine learning that are currently being utilized separately. The physics of AM at all length scales -- solidification, phase kinetics, heat transfer, solid mechanics, etc. -- should be used as first principles for building physics-informed statistical models. Many in the materials science community have considered how to use domain knowledge to build better informatics models \cite{Deaven1995, Morris1996, Wagner2016, Raccuglia2016}. The same should be applied to additive manufacturing.
\end{itemize}

Additive manufacturing stands to significantly expand humanity's ability to manufacture high performance, multifunctional, and highly customized engineering parts. At present, non trivial challenges in understanding the PSPP relationships stand in the way of achieving the full potential of AM. The development, integration, and application of statistical analysis, machine learning, and data-driven approaches into the additive manufacturing R\&D ecosystem will tackle many of the problems currently facing the technology's advancement. Additive manufacturing is positioned to provide foundational case studies for the adoption of machine learning into physics-based integrated computational materials engineering, largely due to the simultaneous peak in funding for both additive manufacturing and data-driven materials research across the globe. The success machine learning applications in metals additive manufacturing are poised to provide the foundation for a new paradigm in integrated computational materials engineering as a whole. 

\section*{Acknowledgements}
We gratefully acknowledge the support of the Department of Defense Office of Economic Adjustment Defense Manufacturing Industry Resilience Program, Michael Gilroy program manager, grants \#CTGG1 2016-2166 and \#ST1605-19-03. NSJ also gratefully acknowledges the support of the Los Alamos National Laboratory Additive Manufacturing Graduate Fellowship given to the ADAPT center to support his graduate research. Citrine informatics is acknowledged for critiquing this manuscript pre-submission.

\bibliographystyle{aipnum4-1}
\bibliography{APR}
\end{document}